\documentclass {article}
\usepackage{graphicx}
\begin {document}
\title {Morphology of Clusters and Superclusters in 
N-body simulations  of Cosmological Gravitational Clustering}
\author {B.S. Sathyaprakash$^1$, Varun Sahni$^2$, Sergei Shandarin$^3$\\
\small
$^1$Department of Physics and Astronomy, Cardiff University of Wales,
Cardiff, CF2 3YB\\
\small
$^2$Inter-University Centre for Astronomy \& Astrophysics,
Post Bag 4, Ganeshkhind, Pune 411007, India\\
\small
$^3$Department of Physics and Astronomy, University of Kansas,
Lawrence, KS 66045, U.S.A.}
\maketitle
\begin {abstract}

We analyse shapes of overdense regions (clusters and superclusters)
in controlled N-body simulations of gravitational clustering with power
law initial spectra $P(k) \propto k^n$, $n = -3, -2, -1, 0.$ 
At values of the density just above the percolation transition 
the number of distinct (isolated) clusters peaks and we use this
`natural threshold' 
to study the shapes and multiplicity function of clusters \&
superclusters. We find that the extent of both filamentarity and pancakeness
increases as the 
simulation evolves, the former being appreciably larger than the latter at 
virtually all epochs and for all spectra considered by us.
Our results also show that high density regions within very massive clusters/
superclusters are likely to be noticeably filamentary or pancake/ribbon-like
when compared to the less dense regions within these objects.
We make a detailed study of two, moment-based `shape statistics' proposed,
respectively, by Babul \& Starkman (BS) and Luo \& Vishniac (LV) and find 
that both LV and BS have certain built-in limitations:
LV is biased towards oblate
structures and tends to overemphasise this property; neither BS nor LV correctly
describe the shape of strongly curved or topologically nontrivial objects.
For instance, a thin filamentary torus and a ribbon are both described by
BS and LV as being pancakes ! By contrast Shapefinders, a new shape diagnostic
{\it not} constructed from density moments but from Minkowski functionals, 
does not suffer from these 
limitations and appears to faithfully reproduce the shapes of both 
simple and 
topologically complex objects.

\end {abstract}

\onecolumn
\section {INTRODUCTION}

Large angle redshift surveys appear to confirm that, far from being
randomly distributed, galaxies are clustered and appear to form
a cellular/filamentary
network consisting of clusters and superclusters of galaxies which are
separated from each other by large voids. The clustering properties of
galaxies have been successfully studied by means of statistical
indicators such as the two-point correlation function, the 
probability density function, etc. (\cite{sc95}).
Useful as these indicators are, they say
very little about the large scale texture of the galaxy distribution
or its morphology. 
Such questions are meaningful since, as demonstrated
in a variety of numerical experiments, galaxies can mesh together in
a number of distinct ways, which could be suggestive of a ``meatball''
topology, ``sponge'' topology, ``bubble'' topology, etc. 
(\cite{delgh91,gmd86,gwm87,mel90,ysf97}).
In addition, the
presence of spectacular large scale features in  galaxy catalogues such
as ``great walls'' in the combined CfA-SSRS surveys, begs the question
as to whether such features are one-dimensional ``filaments'' or two-dimensional
``pancakes/ribbons''.

The shapes (and topologies) of individual clustered objects are important
quantities which are bound to shed light on physical processes responsible
for galaxy clustering. In the so-called `standard model', wherein
clustering is driven by gravitational instability 
and the initial perturbations are assumed to be smooth, conventional
wisdom is that the first nonlinear objects to form are likely to be
`pancakes' (Zel'dovich 1970, Shandarin et al. 1995).
However, as recent work has emphasised, 
the density distribution comes to be dominated by filaments which
weave nearby clusters into a `cosmic web' (Bond, Kofman \& Pogosyan 1996). 
There is no contradiction between these statements. They simply stress
different aspects of nonlinear distributions. The former simply 
describes the
shape of the first caustics. The latter stresses the fact that in a smoothed
density field (caustics must not be present) the percolation threshold 
from disconnected clumps to
connected structure (in other words from the meatball to sponge topology),
occurs at a higher density threshold than
the other percolation transition from the sponge topology to the bubble
topology. This is correct for every generic field. 
As gravitational
clustering advances, filamentarity increases and planarity too remains
statistically significant (Sathyaprakash, Sahni \& Shandarin 1996).
Thus, the Universe
during late stages of gravitational clustering (hierarchical clustering 
or hierarchical pancaking, see e.g., \cite{cmsh93})
consists of a variety of geometrical objects, whose shape and relative
abundance are likely to depend both upon the initial spectrum of
density perturbations and upon the epoch at which clustering is being
studied. 
Other scenario's of structure formation in which large
scale structure is formed by cosmological explosions or seeded by
topological defects (strings, textures) will undoubtedly predict a
different evolution for clustering and hence also a different morphology
for the clustered objects.

Strictly speaking, idealised Zel'dovich pancakes form locally, and whether or
not coherence exists over large scales so that the first objects 
form as large sheets of matter, depends largely upon the primordial fluctuation
spectrum.
Moreover, since Zel'dovich pancakes are transient features which arise
when first caustics form, it is not likely that matter will still be organised
in two-dimensions during the later, strongly non-linear stages of 
gravitational instability. 
Finally, although formally the density in a
Zel'dovich pancake is infinite, in a realistic situation pancakes
are expected to be of finite density --- less dense than filaments since
collapse is along a single dimension for pancakes, whereas
filaments form by the two-dimensional
collapse of matter (Shandarin \& Zeldovich 1989).
Thus, filaments are likely to be more prominent than pancakes, as has 
indeed been observed and quantified for N-body simulations
(Klypin \& Shandarin 1993; Sathyaprakash et al. 1996).
The above examples illustrate that it is 
necessary to address the evolving morphology of large scale structure 
by using good statistical indicators.  Simple methods, such as
fitting an ellipsoid to a cluster of galaxies, could lead to an erroneous
picture, since structures found in 
surveys need not have any
definite archetypal structure implicitly assumed in such methods. Also,
the morphology of structure often depends on scale. Therefore, one needs
more refined scale dependent statistical tools with which to study morphology.

In this study, we analyse the morphology of overdense regions
arising as a result of gravitational instability using some recently
suggested statistical tools.
We study the three-dimensional `structure functions'
introduced by Babul and Starkman (1992) and also
consider an alternate shape statistic proposed by Luo and Vishniac (1995).
We apply these discriminators to study overdense regions 
(clusters \& superclusters) in scale-invariant
N-body simulations for varying epochs of nonlinearity and different
initial spectra.
The present work extends previous work on the subject
including that of Dubinsky (1992) who used ellipsoid-fitting to study
the shapes of individual density peaks as they evolved under gravitational
instability. A one-dimensional shape statistic originally proposed by Vishniac
(1986) has been
used to study the growth of filamentarity in two-dimensional simulations
of gravitational clustering by Nusser \& Dekel (1990) and Sathyaprakash et. al.
(1995).
More recently, a shape statistic based on spatial  
links between high density regions was discussed in \cite{dave}; 
Sahni,
Sathyaprakash \& Shandarin (1998) proposed a new shape diagnostic based on 
Minkowski functionals.
Our present analysis also extends work by Sathyaprakash et. al. (1996)
to later nonlinear epochs
using a wider range of initial conditions and more robust
and sensitive shape statistics. Our results indicate that:
(i) the shape of overdense regions evolves with time,
(ii) with the development of nonlinearity
both filamentary and pancake-like properties of individual clusters are
enhanced and (iii) individual objects get progressively more 
filamentary than planar as the simulation evolves.
These general results hold for a wide range of initial conditions 
and appear to be generic features of structure formation scenario's 
driven by gravitational instability.  However, quantitatively
speaking, though all spectra initially produce small filaments and pancakes,
at later epochs spectra with steeper (negative) slope produce
larger filaments and pancakes, the former being more prominent.
In this study we shall not confine ourselves to studying peaks of the
density field, rather, we shall
apply shape statistics to {\it all} regions above a given density threshold
and study shapes of isolated objects as well as the  morphology
of large scale structure as a whole.
We shall use percolation analysis to aid us in determining
a suitable density threshold above which to identify clusters.

\section {N-BODY SIMULATIONS} \label {sec:nbody}

The N-body simulations used in our study are produced by a staggered 
particle-mesh code with 128$^3$ particles on a 128$^3$  mesh and a 
corresponding Nyquist wavenumber, $k_{\rm Ny} = 64.$
The initial conditions are generated by the Zel'dovich approximation 
(Klypin \& Shandarin 1983 )  such that the initial power spectrum is a 
simple power-law covering the range $n = 0 , -1$, $-2$ and $-3$.
The models are allowed to evolve 
gravitationally until nonlinear effects change the slope of the power 
spectrum.
This change indicates that phase correlations have developed between the 
originally random initial phases. The extent of nonlinearity can be 
characterised
by the parameter  $k_{\rm nl},$ defined by the equation
$\sigma_\delta=a^2\int_0^{k_{\rm nl}} P(k)d^3k=1,$ where
$P(k)$ is the power spectrum of fluctuations. In this study,
we evolve the simulations to values of $k_{\rm nl} =  
64, 32, 16, 8,$ and $4.$  
The value of $k_{\rm nl}$ relates to the scale of structure formation 
in real space. For a detailed discussion of the N-body    
simulations see Melott \& Shandarin (1993).

In our study of shapes of clusters we only use density fields.
Density fields are derived from the above simulations by a cloud-in-cell 
method, whereby each particle's mass is proportionately spread over a 
$2^3$ cell-volume and rescaled 
(8:1) to produce a $64^3$  density field.  This method implies some 
smoothing at small scales but reduces shot noice so that further smoothing 
is not needed before any analysis.
An ensemble family of four realisations is produced from each combination 
of $n$ and $k_{\rm nl}$ to give assessments of the one-$\sigma$ level 
dispersion for each physical quantity measured.

It is likely that no single model studied can pretend to explain the 
real universe.  We consider them to be toy models that serve as a
good test-bed for our shape statistics.  However, if one 
wishes to get a rough idea of how they may 
relate to the real world we provide the following normalisations. 
We assume that the rms fluctuation in number of galaxies is about unity 
within spheres of  radius  $8 h^{-1}$~Mpc,
the rms mass density fluctuation $\sigma_m$ is parameterised by 
the ``bias  factor'',  $b,$  such that $\sigma_g=b\sigma_m$.  
We shall  assume that $b \simeq 1$ which  is  an  adequate  
assumption  for   these  crude
estimates.  Melott  \&  Shandarin  (1993)  showed that, for the models in 
question, the scale of nonlinearity, measured by the top-hat smoothing 
filter, $R_{\rm TH},$  is approximately two times
greater than $k_{\rm nl}^{-1},$  calculated from the extrapolation of 
linear  theory  (more accurately:  
$R_{\rm TH}\simeq 1.8 k_{\rm nl}^{-1}$ in the  $n= 0, -1$ 
models  and  $R_{\rm TH} \simeq 2.8  k_{\rm nl}^{-1}$
in the $n = -2$  model). Thus, identifying every stage with the
present time one can roughly estimate the
size of a mesh cell: $l_c\simeq 26 h^{-1},$ $12.6h^{-1},$
$6.3h^{-1},$ $3.1h^{-1}$ and $1.6 h^{-1}$~Mpc for $k_{\rm nl}=64, 32, 16,
8$ and $4$, respectively.  In our models the smoothing has been performed
with a top-hat filter having a cubic rather than spherical shape which may
add an additional factor of $(4\pi/3)^{-1/3}=0.6$ (assuming volumes of the
filters are similar: $l_c^3=(4\pi/3)(R_{\rm TH}^{(s)})^3).$ Therefore,
one can view each stage of evolution of the models as the density
distribution seen after smoothing with a top-hat filter of radius
$R_{\rm TH}^{(s)}\simeq 16h^{-1},~8h^{-1},~4h^{-1},~2h^{-1},$ and
$1h^{-1}$ Mpc within box-volumes of
$(64l_c)^3\simeq (1600h^{-1}{\rm Mpc})^3,$ $(800h^{-1}{\rm Mpc})^3,$
$(400h^{-1}{\rm Mpc})^3,$ $(200h^{-1}{\rm Mpc})^3$ and 
$(100h^{-1}{\rm Mpc})^3$, for $k_{\rm nl}=64,$ 32, 16, 8 and 4,
respectively.

The purpose of these estimates is to give a rough idea of the range
of parameters characterising models, and therefore more elaborate calculations
are probably not needed. (More realistic models will be discussed in
Sathyaprakash, Sahni, Shandarin \& Ryu 1998.)

\section {METHODS OF ANALYSIS}

In this Section we briefly describe the statistical tools employed by us 
to characterise
the morphology of large scale structure (LSS). 
We introduce the Babul \& Starkman and Luo \& Vishniac
shape statistics and also describe in detail our method of identifying individual
clusters at a density threshold prescribed by percolation theory.

\subsection {Shape Statistics}

In order to quantify the morphology of structure, both of isolated
clusters and of LSS as a whole, we use suitably
modified versions of the structure functions introduced by Babul
\& Starkman (1992) as also the shape statistics of Luo \&
Vishniac (1995). These authors define the morphology of the distribution
of particles by constructing the first- and
second-moments of the particle distribution around a given point.
Geometrically, the shape and structure functions are nonlinear transformations
of the three eigenvalues of the ellipsoid ``fitted'' to the distribution
of particles around the given point. These transformations are implemented
so as to suitably tune and normalise the parameters describing
morphology
and to factor out the size of the distribution, which does not play a role
in describing `shape'.  Both `shape statistics' and `structure functions'
are described by a doublet consisting of two numbers, 
the first describing 
filamentarity (or prolateness)
and the second characterising planarity (or oblateness).
(Babul and Starkman do use a triad, however the presence
of a single non-linear functional relationship between members of the triad
makes only two of them independent.)
It is interesting that, for an object 
chosen randomly from our N-body simulations, 
both shape statistics do not usually give identical results, 
a feature to which we shall return in later
sections !
 
As we shall demonstrate,
these measures of filamentarity
and planarity do not always reflect the visual shape of an object.
This may be because, although the nomenclature
filaments/pancakes is perfectly valid for extreme cases
e.g. matter spread along an infinitely thin line or in an
infinitely thin plane, one has to be careful in interpreting the values measured by these moment-based shape statistics when we are dealing with objects which
are neither prefectly planar nor perfectly linear. 
The fact that the moments of a distribution do depend on the choice of 
origin makes things even more complicated and
we shall have more to say about this and related issues when we discuss
shape eikonals in Section \ref{sec:shape.eikonals} We shall see there
that the Luo-Vishniac shape statistics has the drawback that it is biased
towards planarity and is therefore not a good statistical measure of
morphology despite being derived in a mathematically rigorous and
appealing manner.

\subsection {Global and Local Morphology}

Conventionally, one studies the morphology of LSS by placing
windows of increasingly larger size centered on randomly chosen points and
averaging over many such sample points.
However, a certain amount of control should
be exercised in the choice of sample points so as to reduce the inevitable
scatter in the signal. Since our interest is mainly
to study the morphology of galaxy clusters and superclusters
\footnote{We address the issue of morphology of cosmic voids in
a future work.} we can achieve this by choosing only those random points where
the density is larger than a preset threshold. Such a study
of average morphology is very useful in establishing an overall picture.
However, in order for the shape statistics to be useful in discriminating
models of structure formation and to address issues relating to the physics of
gravitational instability it is appropriate to study the morphology
of individual objects, the shape distribution function, dependence of
shape on cluster mass and size, etc. In studying the morphology of
individual objects one has several options and care has to be exercised
in making a choice. For instance, we can use either the centre of mass
to be the origin or the geometrical centre; we can enclose the whole
object in a window of size larger than the object or study its shape
as a function of window size starting with a small window centered around
the origin. We could either weigh the shape functions with the local value of the
density or
choose not to weigh them. Each of these methods has its own merit and
yields information complementary to the others.
The study of morphology of individual objects can therefore be very
rewarding and rich in information content.
 
It is also important to study the
morphology of structure at different density thresholds since one and the same
object will generally exhibit different properties at varying thresholds.
For example, at very high density or brightness thresholds a 
barred spiral galaxy
surrounded by a dark halo would look filamentary,
at moderate densities a disc and at very low
densities a spheroid.  Therefore, in this work we not only study the global
morphology of structure but also the morphology of isolated clusters and
superclusters at different density thresholds.

\subsection {Identifying Structures using Percolation Theory} \label{sec:percolation}
The issue of density threshold mentioned earlier can bring forth an
element of subjectivity into the analysis and the resulting picture can be confusing if density thresholds are chosen in an arbitrary manner.
We try to introduce some objectivity in choice of threshold
by using percolation analysis,
which is a tool to understand connectivity of structure and hence is
closely linked to the study of morphology.
 
Our method of identifying structures is the following: at any given
density threshold a cluster is defined 
using a `friends-of-friends' algorithm with
six nearest neighbours. At very high thresholds only a small volume
is in the overdense phase, consequently there are very few clusters and the
largest of them is very small. As the density
threshold is lowered the overdense phase grows, leading to an increase in
the number of clusters and a concomitant growth in the volume (and mass) of the
largest cluster. Indeed, lowering the density contrast further causes many
small clusters to merge with the largest cluster, leading to a sharp increase
in the volume of the latter until, at the percolation transition,
the largest cluster
runs across the entire region of interest 
and has a volume  which is a significant fraction $(> 10\%$)
of the total volume in the overdense
phase. 
It turns out that the
number of distinct clusters is most abundant {\it just before} the
percolation transition which leads us to feel that the percolation threshold
provides
a natural choice at which to identify structures (see Fig.~\ref {fig:perc}).
We also use other thresholds in addition to percolation 
(see below) at which to identify clusters, 
demonstrating in the process, that the main conclusions of this study are not
sensitive to choice of threshold. (Other aspects of percolation theory are
discussed in \cite{ys96,sss97a}.)
  
We now introduce the shape statistics used
in our study. 

\subsection {The Babul-Starkman Statistic}

Let $\vec x^p = (x^p_1, x^p_2, x^p_3),$ $p=1,\ldots,N,$
denote the position vector of the $p$th particle in a system consisting of 
$N$ particles. The first- and second-moments of the particle distribution 
around a fiducial point $\vec x^0$ are given by
\begin {eqnarray}
M_i(\vec x^0; R) & = & \frac {1}{N} \sum_{p=1}^N
\left (x^p_i-x^0_i \right ) W\left (|\vec x^p-\vec x^0| \right ),\nonumber \\
M_{ij}(\vec x^0; R) & = & \frac {1}{N} \sum_{p=1}^N
\left (x^p_i-x^0_i \right ) \left (x^p_j-x^0_j \right )
W \left (|\vec x^p-\vec x^0| \right ),\end {eqnarray}
where $i,j=1,2,3$ and $W$ is a window of effective radius $R$
centred on the point $\vec x^0$. 
The moments do, in general, depend
on the nature of the window function. In this study, we use
a simple spherical window function.
Given the moments of the distribution one can compute the
moment of inertia tensor $I_{ij}$ 
\begin {equation}
I_{ij} = M_{ij} - M_i M_j.
\label {eq:bs0}
\end {equation}
The three eigenvalues $I_1,$ $I_2$ and $I_3,$ of the inertia
tensor represent
the three principal axes of an ellipsoid fitted to the matter distribution
around the point in question, they therefore contain information
about how the matter is distributed around a given point. (This statement is
rigorously true if $R$ is infinitesimally small making it possible for the
density field about a local maximum 
to be  approximated by a quadratic, which in turn, is equivalent to fitting by an 
ellipsoid.) The absolute value of the eigenvalues is
not important in quantifying morphology and therefore it
suffices to consider their ratios
\begin {equation}
\mu \equiv \left (\frac {I_2}{I_1} \right )^{1/2}, \ \ 
\nu \equiv \left (\frac {I_3}{I_1} \right )^{1/2},
\end {equation}
where we assume that eigenvalues are arranged 
in order of increasing magnitude, i.e., $I_1\le I_2\le I_3.$ 
The shape statistic of Babul \& Starkman (1992) (henceforth BS)
consists of a triad of numbers $(S_1, S_2, S_3)$ and 
is constructed out of the parameters $\mu$ and $\nu$ as follows: 
\begin {eqnarray}
S_1 & = & \sin \left [ \frac {\pi}{2} \left (1-\mu \right)^p \right ],\nonumber\\
S_2 & = & \sin \left [ \frac {\pi}{2} a(\mu, \nu) \right ],\nonumber\\
S_3 & = & \sin \left [ \frac {\pi}{2} \nu \right ],
\label {eq:bs1}
\end {eqnarray}
where $p=\log 3/\log 1.5,$ the function $a(\mu,\nu)$ is implicitly
given by
\begin {equation}
\frac {\mu^2}{a^2} - \frac {\nu^2}{a^2(1-\alpha a^{1/3} + \beta a^{2/3})}
= 1,
\label {eq:bs2}
\end {equation}
and the parameters $\alpha$ and $\beta$ are defined by
\begin {equation}
\alpha =  1.9, ~\beta = - \frac {7}{8} 9^{1/3} + \alpha 3^{1/3}.
\label {eq:bs3}
\end {equation}
These definitions ensure $0\le S_i \le 1$, $i = 1,2,3$ and $|\vec S| \le 1$ for 
the `shape vector' $\vec S= (S_1, S_2, S_3),$ 
which, therefore, always lies in the first octant of a 
three-dimensional `shape-space'.

The set of all shape vectors (points associated with the tip of
the shape vector) spans a three dimensional volume, every point located within
this volume corresponding to a certain geometrical shape.
For a spherical distribution we have $I_1=I_2=I_3$ which implies that 
$\mu=\nu=1$ and $a=0$ so that $(S_1, S_2, S_3)= (0,0,1).$ 
For a planar distribution $I_1=I_2,$ $I_3=0$ which implies 
$\mu=1,$ $\nu=0$ and $a=1$ so that $(S_1,S_2,S_3) = (0,1,0).$ 
Finally, for a linear distribution $I_2=I_3=0$ so that
$\mu=\nu=a=0$ and $(S_1,S_2,S_3)=(1,0,0).$ Consequently, 
BS call $S_1$ filamentarity, $S_2$ planarity and
$S_3$ sphericity. For a distribution of matter which is neither perfectly
spherical nor planar nor linear: $S_i\ne 0$, $i = 1,2,3.$
In such cases it is best not to use any qualifying words to
describe structure but speak in terms of the magnitude and orientation
of the shape vector. From (\ref{eq:bs1}) -- (\ref{eq:bs3}) one 
finds that $S_1$, $S_2$, $S_3$ are not independent, a knowledge of
any two of them effectively determines the third.
We shall work with $S_1$ and $S_2$ whose maximum
values characterise a filament and a pancake respectively; 
small values of both $S_1$ and $S_2$ imply a large sphericity ($S_3$) for the
distribution.

Clearly, the moments as defined above are only useful for point processes
and cannot be used if we want to address the morphology
of continuous distributions such as density fields, surface brightness, etc. 
We therefore need to generalise these concepts
to fields defined in space.
This can easily be accomplished by defining
density-weighted moments as follows
\begin {eqnarray}
M_i(\vec x^0; R) & = & \frac {1}{\cal M} \sum_{p=1}^N
\left (x^p_i-x^0_i \right ) \rho\left (\vec x^p \right )
W\left (|\vec x^p-\vec x^0| \right ),\nonumber \\ 
M_{ij}(\vec x^0; R) & = & \frac {1}{\cal M} \sum_{p=1}^N
\left (x^p_i-x^0_i \right ) \left (x^p_j-x^0_j \right ) 
\rho\left (\vec x^p \right )
W\left (|\vec x^p-\vec x^0| \right ),
\label {eq:bs4}
\end {eqnarray}
where $i,j=1,2,3,$ and $\cal M$ is the total mass in the region of interest:
\begin {equation}
{\cal M} = \sum_{p=1}^N \rho\left (\vec x^p \right ) 
W\left (|\vec x^p-\vec x^0| \right ).
\label {eq:bs5}
\end {equation}
substituting (\ref{eq:bs4}) in (\ref{eq:bs0}) -- (\ref{eq:bs1}) we obtain 
`structure functions' $\vec S = (S_1, S_2, S_3)$ defined for a continuous density field.

Since $\rho$ could, in principle, be an arbitrary scalar quantity
it is clear that the shape statistic
can now be computed for any scalar field, be it a density field,
surface brightness, temperature field, etc., this useful modification allows
us to apply shape statistics to a wider variety of situations.

\subsection {The Luo-Vishniac Statistic}

The BS statistic is derived heuristically and (we guess) by experimenting
with the behaviour of the shape vector as different parameters, such
as $p,$ $\alpha$ and $\beta$ are varied. Luo and Vishniac 
follow a more rigorous approach. The starting point for their shape statistic is
once again the moments of the distribution. They introduce a shape
function which is the most general and coordinate-independent 
linear combination of the moments of cubic non-linearity:
\begin {eqnarray}
S(R) & = &\frac {1}{M^3} \biggl [ 
a M \left (M_{ij}-M_iM_j \right)
    \left (M_{ij}-M_iM_j \right) 
+ b M \left (M-M_iM_i \right)^2 + \nonumber \biggr .\\
& & \biggl . c M_{ij} \left (M_{ik}-M_iM_k \right)
    \left (M_{jk}-M_jM_k \right)
+ d M_{ij} \left (M_{ij}-M_iM_j \right)
    \left (M-M_k M_k \right) \biggr ],
\end {eqnarray}
where a summation over repeated indices is implied.
As earlier $R$ is the radius of the window function, $M=M_{ii}$
is the trace of the second-moment, and $a$, $b,$ $c,$ and $d$ are
constants.  The constants
$a,$ $b,$ $c$ and $d$ can be determined by demanding that the 
geometrical construct $S(R)$ be consistent with what we intuitively expect for the
morphology of archetypal point distributions. By requiring that the shape
function vanish for a uniform distribution of particles within the
region of interest one obtains two constraints
which fix two out of the four constants.
In order to fix the remaining two one can either demand that the shape function
be unity for an infinitely thin linear distribution and
zero for an infinitely thin planar distribution or vice versa.
In the first case one obtains the statistic $S_l$ 
and in the second case one obtains $S_p,$
which Luo and Vishniac call linearity and planarity, respectively:
\begin {equation}
S_l \equiv S(R), a=-1,\ b=1/2,\ c=3,\ d=-3/2
\end {equation}
\begin {equation}
S_p \equiv S(R), a=4,\ b=-4,\ c=-12,\ d=12.
\end {equation}
Thus, the Luo-Vishniac statistic (henceforth LV)
consists of a doublet of numbers $S_l$ and
$S_p$ both of which satisfy $0 \le S_l, S_p \le 1.$

Having introduced the BS and LV shape functions we now proceed to test their
effectiveness in determining shapes.

\section {Shape Eikonals} \label {sec:shape.eikonals}

In this Section we address the following questions related to 
the internal consistency and robustness of the
shape statistics: (i) is the statistic a good measure of morphology
and consistent with the visual impression conveyed by a
structure? (ii) is the statistic unbiased? The important related issue of 
whether the statistic performs reasonably well in the presence of  
noise, has been amply addressed by BS and LV
who concluded that the statistic they had proposed 
was reasonably robust.
We return to this issue in a future work when we shall address the performance of 
different shape statistics 
in the presence of noise (\cite{bsss98}). For the present we investigate whether 
the BS and LV shape statistics correctly diagnose the shape of an object. 

Our first aim is to test the performance of shape statistics
using different eikonal shapes for this purpose.
(i) We study the behaviour of the shape
statistics as a sphere is continuously deformed into an oblate or 
prolate ellipsoid and describe this behaviour  using
trajectories in
`shape phase-space' representing continuous deformations; (ii) we study the
morphology of objects with a non-trivial topology, such as
a torus, this study leads to an important conclusion, namely, 
descriptions of morphology (of LSS) are incomplete if 
only moment-based shape statistics are used, the latter must be supplemented
with shape diagnostics sensitive to topology
(Sahni, Sathyaprakash \& Shandarin 1998); 
(iii)  we randomly choose isolated structures (clusters/superclusters)
from an N-body simulation and apply the BS and LV statistics to study their 
shapes. This permits us to say something about the extent of the shape 
phase-space sampled, respectively, by BS and LV. 

\subsection {Structural Transformations and `Shape-Space'}

As mentioned in the last Section, the doublets $(S_l,~S_p)$ and $(S_1,~S_2)$ 
can be treated as two-dimensional vectors with norm $\le 1$. 
The following questions then arise:
given an ensemble of geometrical objects 
what possible values could these vectors take? In other words
what is the {\em shape phase-space} sampled by the two statistics?
To explore this in a heuristic way\footnote{A rigorous answer to this
question is rather complicated and unnecessary for our purposes; 
such a proof would need a more mathematical treatment of the subject than
is dealt with here.} we consider geometrical transformations
of a sphere into oblate and prolate ellipsoids (\cite{bs92}).
(Henceforth we shall refer to {\em shape phase-space} as simply 
{\em shape-space}.)

\begin {figure}[ht]
\centering
\begin {minipage}[c]{3in}
      \includegraphics[width=3in]{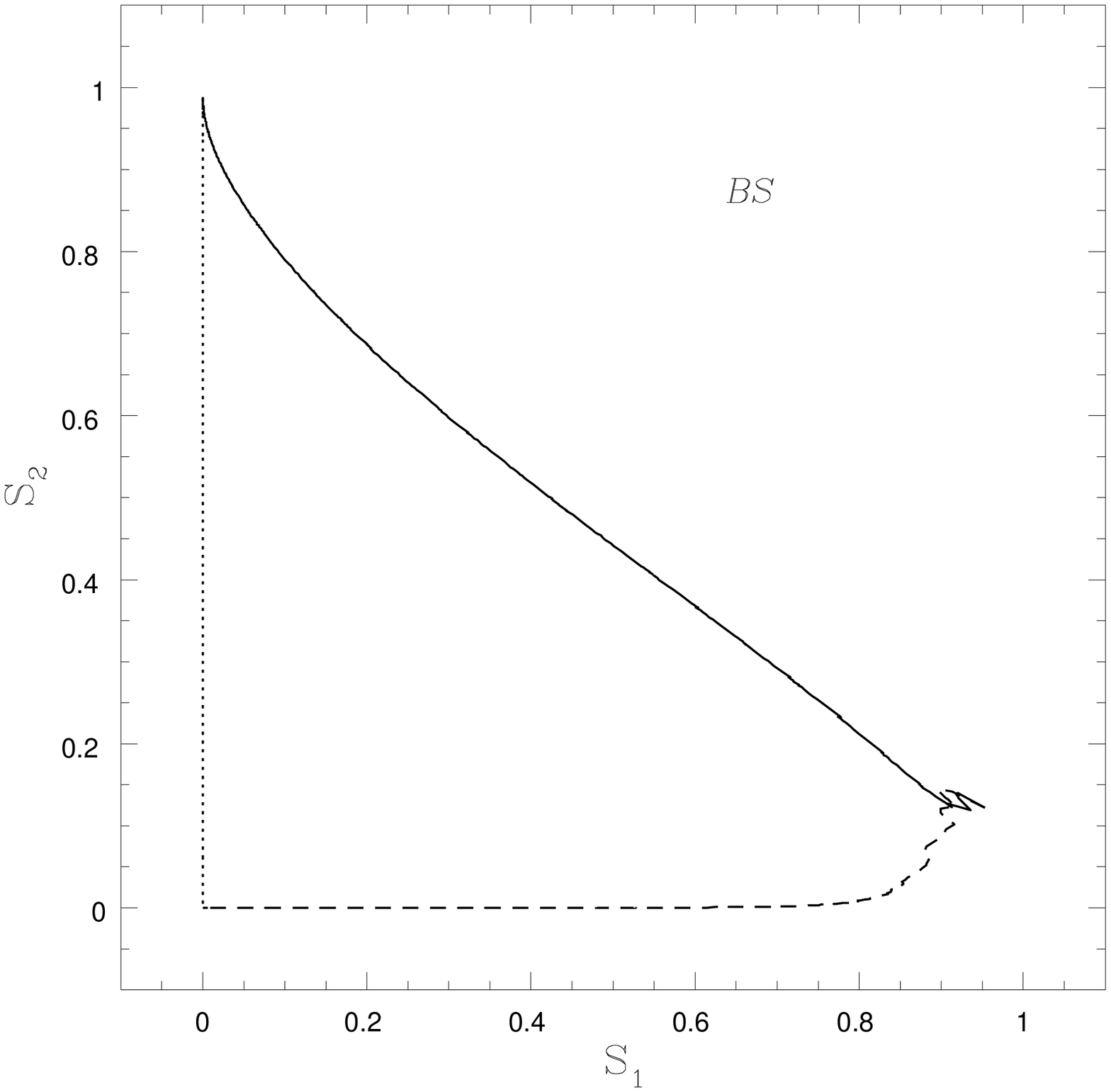}
\end {minipage}
\begin {minipage}[c]{3in}
      \includegraphics[width=3in]{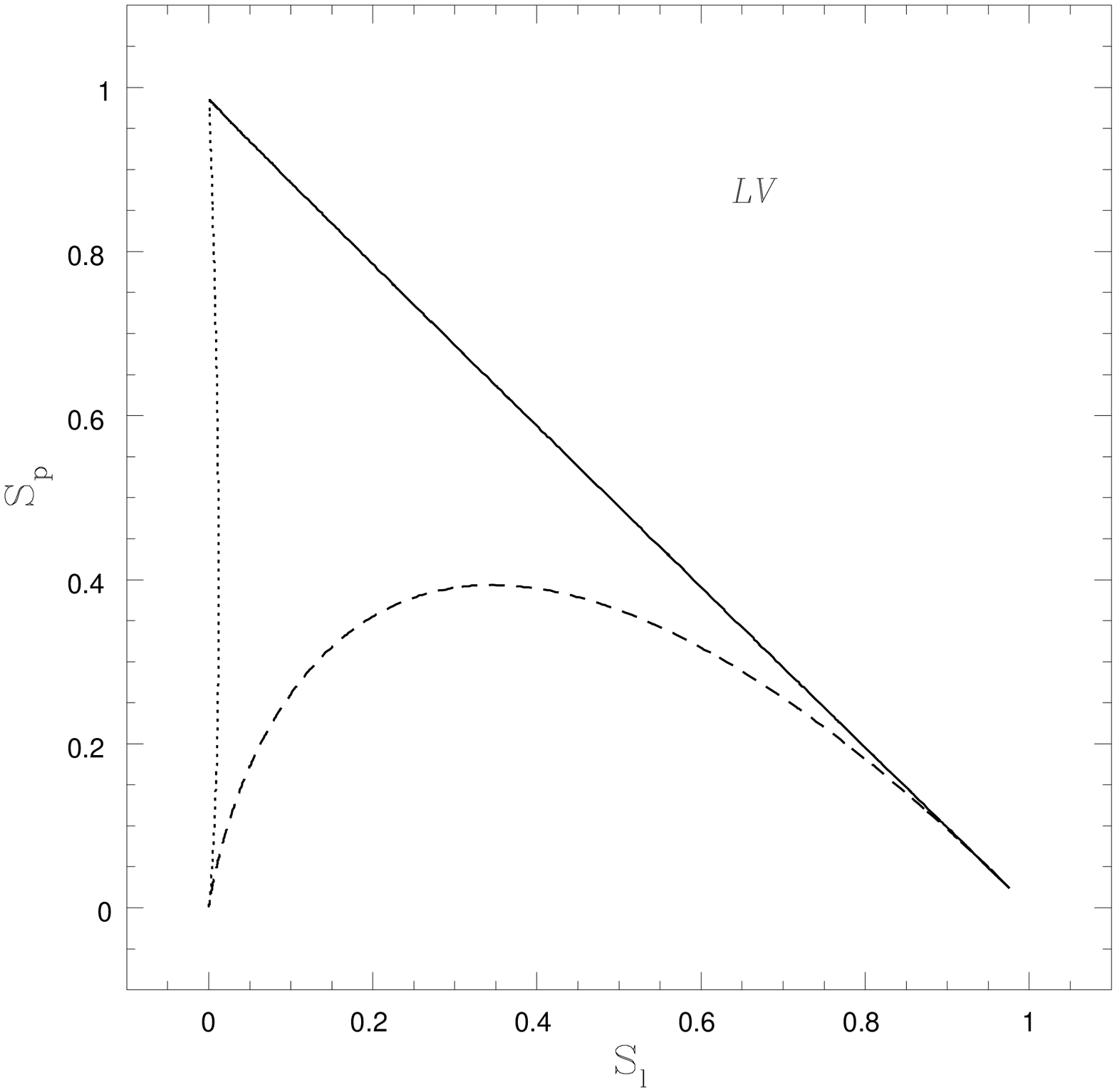}
\end {minipage}
\caption {Structural deformations of a sphere: 
sphere $\rightarrow$ pancake (dotted line), 
pancake $\rightarrow$ filament (solid line) and
filament $\rightarrow$ sphere (dashed line), are illustrated in the form
of a `shape phase-space' for the Babul-Starkman statistics ($S_2$ versus
$S_1$, left panel) and Luo-Vishniac statistics 
($S_p$ versus $S_l$, right panel).}

\label {fig:phase.space}
\end {figure}

Let $r_x \ge r_y \ge r_z$ denote the three principal radii of an
ellipsoid. Since the size of the object is unimportant let us fix
one of the radii, say $r_x,$ to unity.
In Fig.~\ref {fig:phase.space} we have plotted $S_2$ vs $S_1$ and
$S_p$ vs $S_l,$ as a sphere ($r_x=r_y=r_z=1$) with
$(S_1, S_2)=(S_l, S_p)=(0,0)$, is continuously deformed either into an oblate
ellipsoid ($r_x=r_y=1, 0\le r_z \le 1$) ending up in a plane ($r_z=0$)
with $(S_1, S_2)=(S_l,S_p)=(0,1);$ or into a
prolate spheroid ($r_x=1,r_y=r_z,$ $0\le  r_y,r_z \le 1$) ending in a line 
($r_y=r_z=0$), with $(S_1, S_2) = (S_l, S_p) = (1,0).$ 

First consider the BS statistic.
As the sphere is continuously deformed into a plane, 
$S_2$ (planarity) increases 
from 0 to 1, while $S_1$ (filamentarity) remains zero throughout.  
Similarly, as the sphere is deformed into a line, $S_2$ remains fixed at
zero while $S_1$ increases from 0 to 1 (cf. Fig.~\ref{fig:phase.space} left panel).
(During these deformations $S_3$ (sphericity) 
decreases from 1 to 0.) Such a behaviour has been made possible by a 
proper tuning of parameters
($p,$ $\alpha$ and $\beta$ in Eqs.~(\ref {eq:bs1}) and (\ref{eq:bs2})) in 
the BS statistic (\cite{bs92}). Based on the above `experiment' it is 
reasonable to expect that if objects of all possible shapes are analysed, 
the BS statistic would fill in a triangle in `shape-space'. 
(For both BS and LV statistics shape shape-space is 
triangular since, according to these statistics,
no object can be simultaneously a pancake and a filament. 
However, such objects might occur in nature; for example
a long ribbon has a mixed morphology of both a pancake and a filament -- see 
Table \ref{table:tor}.)

Next, let us consider the LV statistic.
As a sphere is deformed into a plane, we find that while $S_l$ (linearity)
remains zero, $S_p$ (planarity) increases from 0 to 1. 
However, the LV statistic  shows anomalous behaviour as a 
sphere is deformed into a filament. Notice that for this deformation
$S_p$ {\it does not} remain fixed at zero as $S_l$ increases from 0 to 1.
Contrary to expectations planarity initially {\em increases} as a sphere is 
deformed into a filament.
Consequently, shape-space is only {\it partially filled} as seen
in Fig. \ref {fig:phase.space}b.
Thus, according to this statistic, while an object can 
be planar without any attributes of linearity, it {\em cannot} be linear
without having a measure of planarity. 
This anomalous behaviour demonstrates that the LV statistic is biased
towards planarity and may not be a suitable tool with which to 
study the morphology of large scale structure.
This unattractive feature, present in a statistic which 
in all other respects is rigorously defined, 
reminds us of the complexities of constructing a good shape statistic 
and the related difficulty in modelling and quantifying the morphology of
LSS. It might be that a certain amount of parameter `tuning' (as in BS)
is necessary in order to
obtain a good moment-based shape statistic.

\subsection {Structures with non-trivial Topology}\label{sec:torus}

Sometimes what our mathematical constructs measure
may not be consistent with the visual impression conveyed 
by an object. To illustrate this, consider a 
torus having an elliptical cross-section described by the parametric form
\begin {equation}
{\bf{r}} = (b + c\sin{\phi})\cos{\theta}~\hat{x} +
(b + c\sin{\phi})\sin{\theta}~\hat{y} + a(\cos{\phi})\hat{z}
\label{eq:torus}
\end {equation}
$a,c < b$, $0 \le \phi, \theta < 2\pi$.
The torus has diameter $2\pi b$ and an elliptical cross-section,
$a$ and $c$ being, respectively, radii of curvature
perpendicular and parallel to the plane of the
torus. The circular torus is given by $a=c$.
Geometrically, an infinitely thin circular torus is 
one-dimensional while a sheet is two-dimensional.
However, the moment of inertia about the centre of mass
cannot distinguish between the two distributions,
the ratios of eigenvalues of the inertia tensor, $\mu$ and $\nu,$ being
exactly the same for both objects.
As both BS and LV 
are moment based statistics, torus and
sheet are declared to be sheet-like, i.e., we get
$(S_1,S_2)=(S_l,S_p)=(0,1)$ for both objects!

\begin{figure}
\centering
   \begin {minipage}[c]{3in}
      \includegraphics[width=3in]{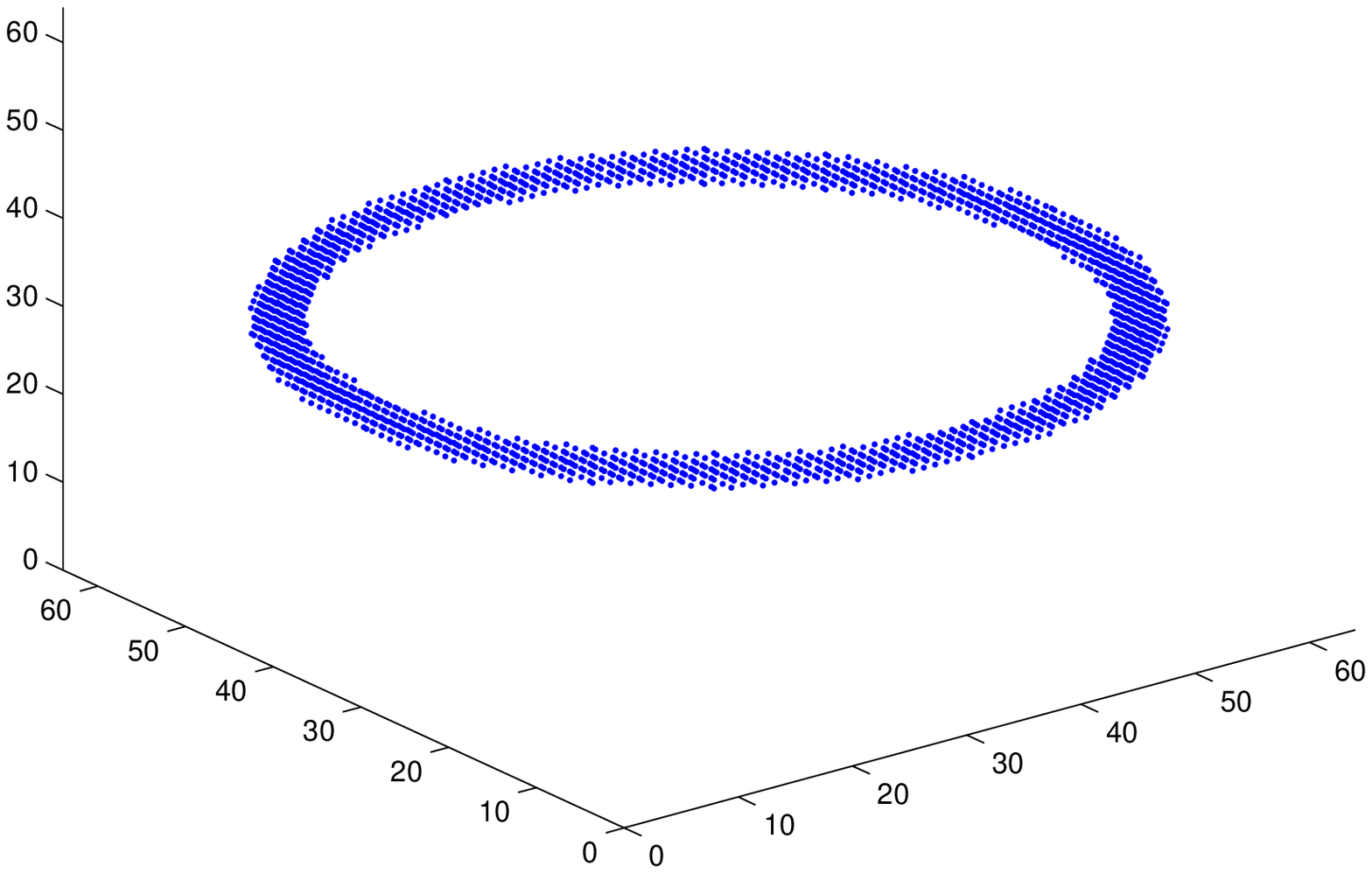}
   \end {minipage}
   \begin {minipage}[c]{3in}
      \includegraphics[width=3in]{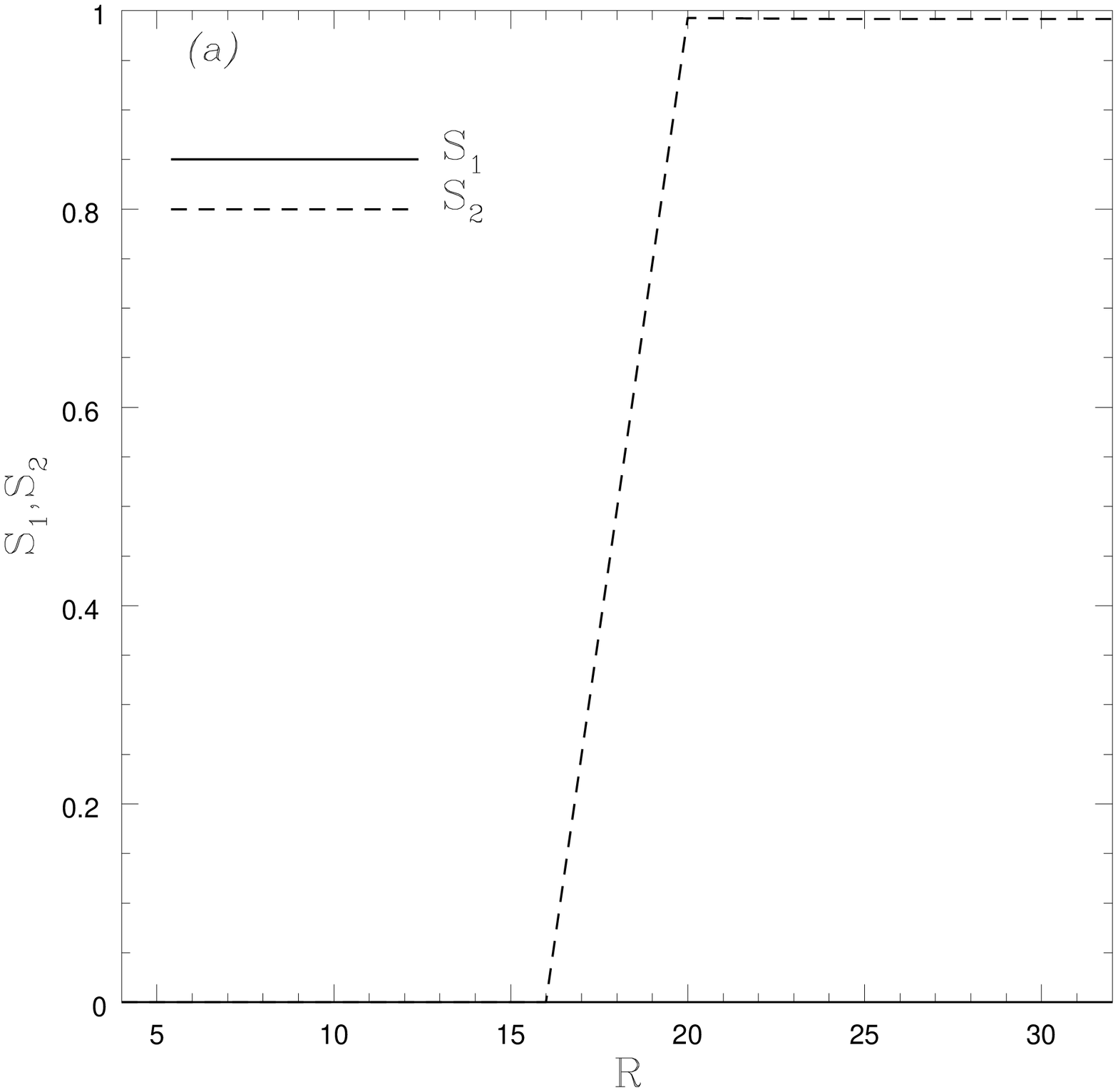}
   \end {minipage}
   \begin {minipage}[c]{3in}
      \includegraphics[width=3in]{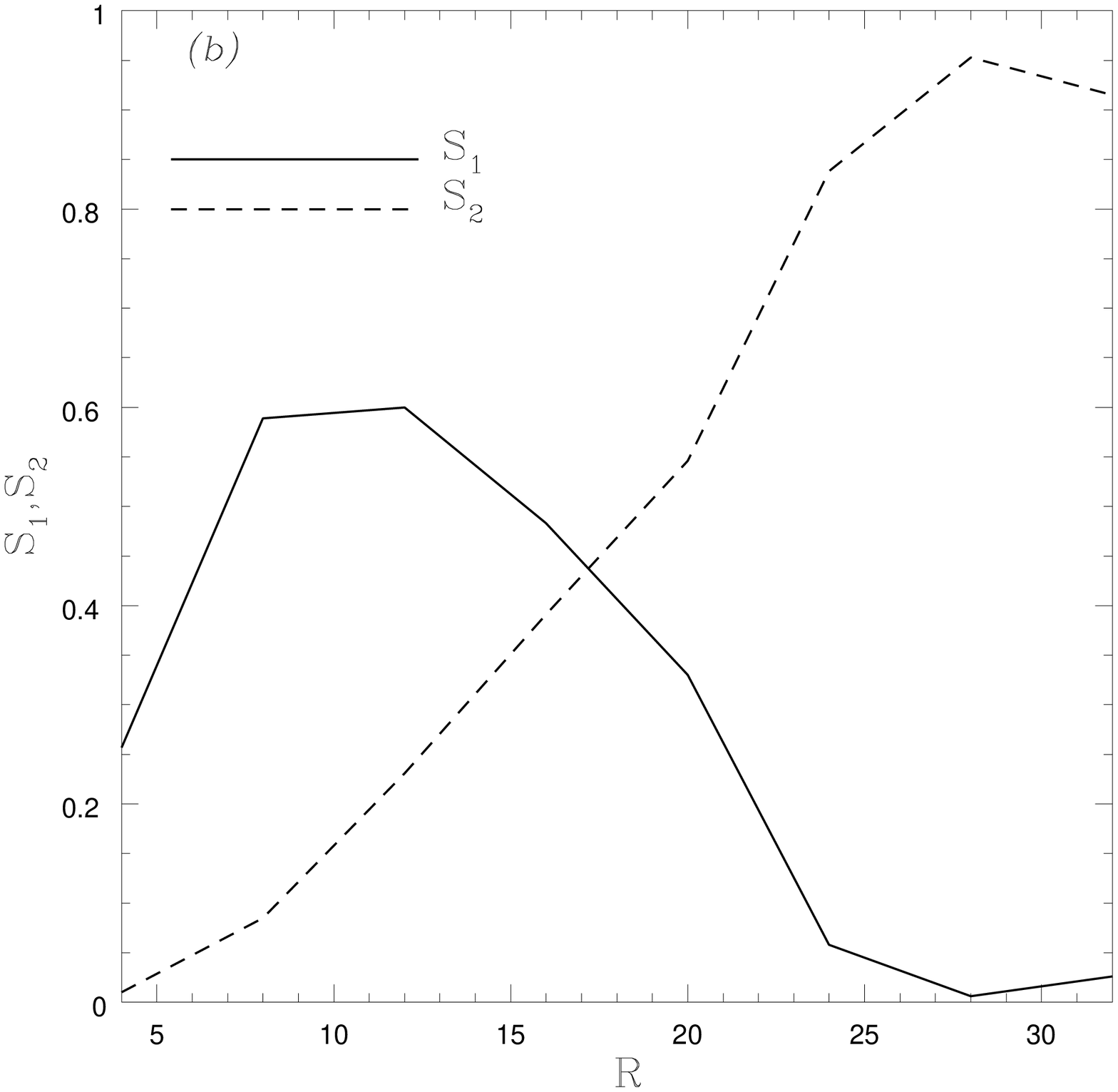}
   \end {minipage}
   \begin {minipage}[c]{3in}
      \includegraphics[width=3in]{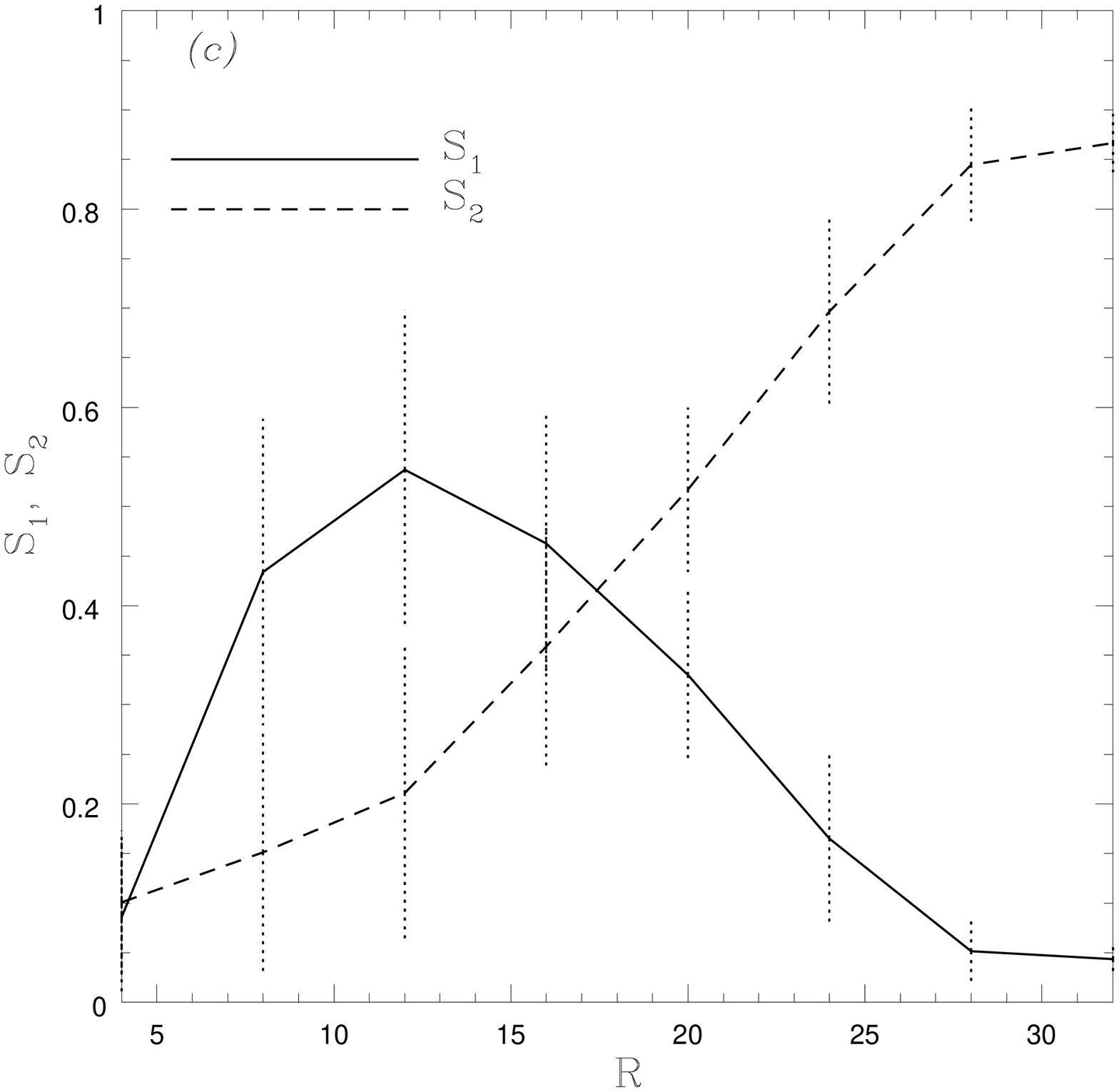}
   \end {minipage}
\caption {BS shape parameters $S_1$ and $S_2$ are shown plotted as a function
of the window size for the toroidal distribution of matter shown in the
upper left panel. 
The window is located at the centre-of-mass in (a) and at 
a point enclosed by the surface of the torus in (b). 
In (c) average values of $S_1$ and $S_2$
are computed using a number of random points.}
\label {fig:torus}
\end{figure}

We therefore see that if we were to study shape statistics
with moments evaluated about the centre of mass,
sometimes we could get a spurious result.  For a torus this arose because
the torus does not enclose its centre of mass within itself.
This problem can be partially alleviated by determining the moments about
a point enclosed by the surface of the torus, rather than about its
centre of mass. If we do this then,
for small sizes of the window, the statistics do declare the torus to be
a filament (i.e. $S_1 \simeq 1$ and $S_2 \simeq S_3 \simeq 0$). However, 
as the window radius gets larger, 
filamentarity reduces and planarity increases, till the torus
is once more declared to be a planar object ($S_1 \simeq S_3 \simeq 0$ 
and $S_2\simeq 1$) when the size of the
window exceeds the radius of the torus. 
Another method which might yield sensible
results
would be to average the shape statistics by choosing a number of different
points in a window that completely encloses the object. This would 
declare the torus to be an intermediate object having both filamentary
and planar properties.
To illustrate these different methods we 
consider a circular torus which is shown 
in the upper-left corner of Fig.~\ref{fig:torus}. The BS statistics 
$S_1$ (solid-line) and $S_2$ (dashed-line) are plotted as functions of the
size $R$ of the window centred at (a) the centre-of-mass 
(upper-right) and (b) a random point on the torus (lower-left). 
We also show averaged BS statistics obtained by choosing (c) a number 
of random centres in the region enclosing the torus (lower-right).
(The LV statistics give similar results and we do not show them
here.)

Of course, it is unlikely that in any simulation or survey one will find
a perfect torus; nor for that matter is one likely to find a perfect filament or
sheet. In cases where the matter distribution is one-dimensional,
but not along a straight line, one is bound to get a somewhat smaller signal for
filamentarity from both BS and LV 
than if the distribution is perfectly straight. This is
even more true for two-dimensional distributions: An infinitesimally thin
spherical shell and a distribution which is isotropic about a given point
have identical
moments about their centres and will be declared `spherical' by both BS and LV.
Similarly, wiggly, wavy or bent two-dimensional surfaces will not be declared 
sheets by these statistics; i.e., $S_2, S_p\ne 1$. 
Moreover, as predicted by the Zeldovich 
approximation and the adhesion model, cosmological pancakes are not 
as prominent as filaments or clumps nor are they likely to be perfectly flat,
planar objects. This makes it difficult for moment-based 
shape-statistics to detect pancakes in N-body simulations and in galaxy
surveys. Problems such as these have prompted us to
explore other statistics to describe morphology which not only take
into account shape but topology of structure as well. 
{\em Shapefinders} -- shape diagnostics constructed from fundamental properties
of a surface, such as its Minkowski functionals 
which we describe below,
is an example of a statistic which can successfully
diagnose shapes of topologically simple and complicated surfaces, without
fitting them with ellipsoids (for details see Sahni, Sathyaprakash and 
Shandarin 1998; Minkowski functionals are discussed in \cite{mbw94}). 

{\it Shapefinders} for a compact surface (which could be an
isodensity surface in simulations or surveys) are constructed from the following 
four `Minkowski functionals' :
(i) Volume $V$, (ii) surface area $S$, (iii) integrated mean curvature:
$C = {1\over 2}\int(\kappa_1 + \kappa_2) dS$,
(iv) integrated Gaussian curvature (genus):
${\cal G} = -{1\over 4\pi}\int \kappa_1\kappa_2 dS,$ where
$\kappa_1 \equiv 1/R_1$, 
$\kappa_2 \equiv 1/R_2$ are the principal curvatures at a point
on the surface.
Multiply-connected surfaces have ${\cal G} \ge 0$ while simply connected
have ${\cal G} < 0$.

The three {\it Shapefinders} having dimensions of [Length] are:
${\cal H}_i, i = 1, 2, 3,$ where
${\cal H}_1 = V/S$, ${\cal H}_2 = S/C$ and ${\cal H}_3 = C$
(for multiply-connected surfaces $C/{\cal G}$ may be more appropriate than
$C$).
We can also
define a pair of {\it dimensionless} Shapefinders
\footnote{Note that our $({\cal K}_f, {\cal K}_p)$ is the same as
$({\cal K}_2, {\cal K}_1)$
in Sahni, Sathyaprakash \& Shandarin (1998). The Shapefinders have been
normalised so that ${\cal H}_i = R~$ $({\cal K}_i = 0$) for a sphere of
radius $R$.}: ${\bf{\cal K}} \equiv ({\cal K}_f, {\cal K}_p)$
\begin{equation}
{\cal K}_p = \frac{{\cal H}_2 - {\cal H}_1}{{\cal H}_2 + {\cal H}_1},\ \
{\cal K}_f = \frac{{\cal H}_3 - {\cal H}_2}{{\cal H}_3 + {\cal H}_2}.
\end {equation}
where  ${\cal K}_{p,f} \le 1$ by construction.
For a pancake (quasi-two-dimensional object which can be curved)
${\cal H}_1 \ll {\cal H}_2 \simeq
{\cal H}_3$ and ${\cal K}_{p} \gg {\cal K}_{f} $; an ideal pancake has
${\bf{\cal K}} \simeq (0, 1)$. 
A filament (a quasi-one-dimensional object, not necessarily straight) has
${\cal H}_1 \simeq {\cal H}_2 \ll {\cal H}_3$ and ${\cal K}_{f} \gg {\cal K}_{p} $;
an ideal filament has ${\bf{\cal K}} \simeq (1, 0)$.
For a sphere 
${\cal H}_1 \simeq {\cal H}_2 \simeq {\cal H}_3$ and
${\bf{\cal K}} \simeq (0, 0)$. An additional surface to consider
is a `ribbon', for which ${\cal H}_1 \ll {\cal H}_2 \ll {\cal H}_3$
and ${\cal K}_{f} \sim {\cal K}_{p} > 0$; an ideal ribbon has
${\bf{\cal K}} \simeq (1, 1)$.

\begin{figure}
\centering
\includegraphics[width=3 true in]{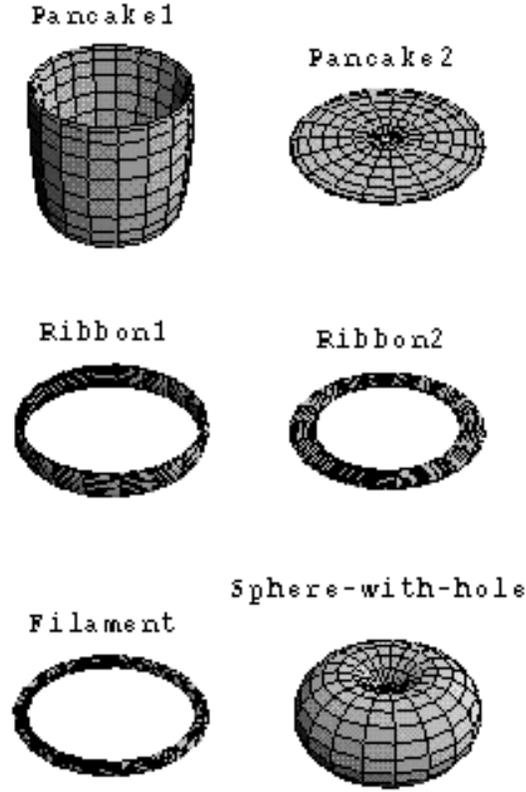}
\caption {Deformations of a torus with elliptical cross-section. The values of
BS, LV and Shapefinders shape statistics for the objects shown here are given in 
Table \protect {\ref{table:tor}}.}
\label {fig:ell_torus}
\end{figure}

In Table \ref{table:tor} we provide values for the dimensionless Shapefinders
${\cal K}_{p,f}$ for deformations of the elliptical torus 
Eq.~(\ref{eq:torus}) shown in Fig.~\ref{fig:ell_torus} and
described in detail in Sahni, Sathyaprakash and Shandarin (1998).
We also give corresponding values of the LV and BS shape statistics.
Our results clearly show that: (i) both LV and BS give low values for 
`pancakeness' (described by $S_p$, $S_2$)
for a hollow circular cylinder, mistaking it for a spherical object !
(ii) a filamentary torus is described by both statistics as a pancake and
(iii) a ribbon is mistaken for a pancake by both LV and BS.
In all cases Shapefinders give a much better description of the shape 
of an object. 

\subsection {Random Sampling of clusters drawn from N-body simulations}

In the last two Sections we have studied the performance of
shape statistics using
well defined `eikonal' shapes. In this section  we study the performance
of BS and LV  using objects drawn at random from N-body simulations.
(A detailed investigation of Shapefinders is presently in progress,
Buchert, Sathyaprakash, Schmalzing, Sahni (1998).)

\begin{figure}
\centering
\begin {minipage}[c]{3in} 
\includegraphics[width=3in]{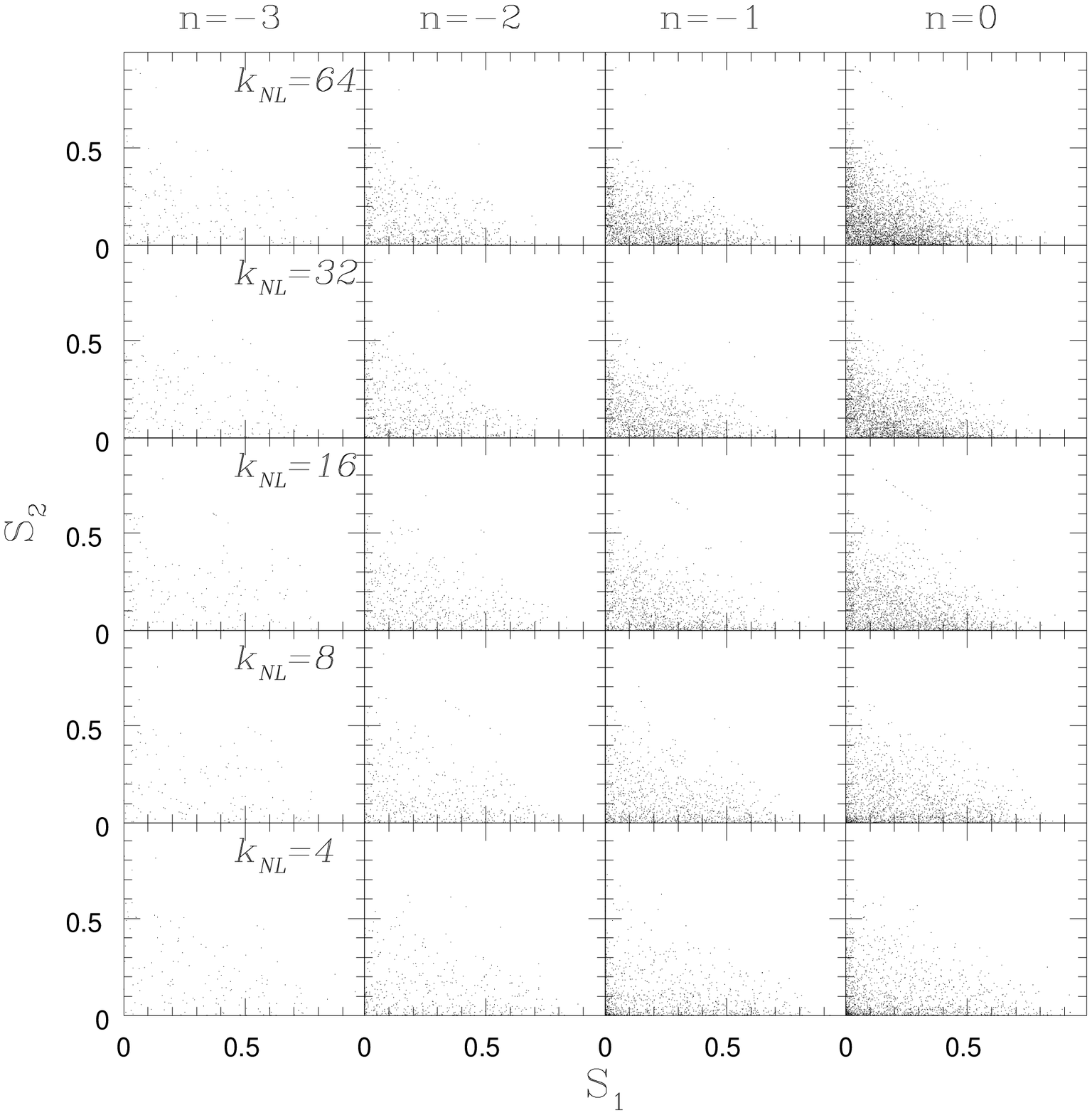}
\end {minipage}
\begin {minipage}[c]{3in} 
\includegraphics[width=3in]{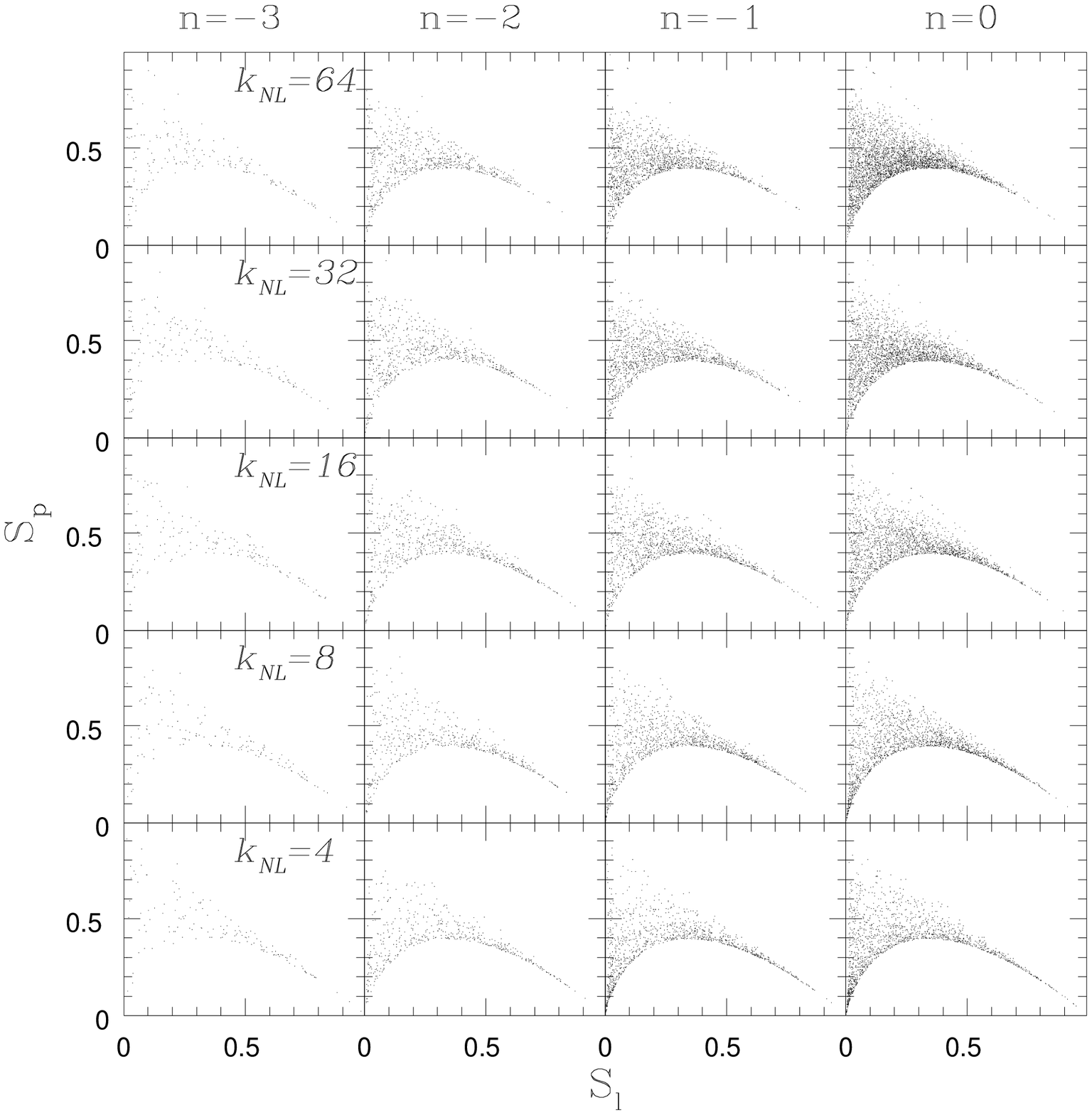}
\end {minipage}
\caption {Scatter plots of (BS) $S_2$ versus $S_1$ (left) and (LV)
$S_p$ versus $S_l$
(right), showing shapes of clusters drawn at random from N-body simulations.}
\label {fig:scatter}
\end{figure}

In Fig.~\ref{fig:scatter} we show scatter plots of
both BS and LV statistics for structures randomly selected from N-body
simulations at the percolation
threshold. Conclusions drawn from the study of eikonal shapes are
borne out in these scatter plots.  BS clearly fills the
entire shape-space whereas LV does not.

\begin{figure}
\centering
\includegraphics[width=3in]{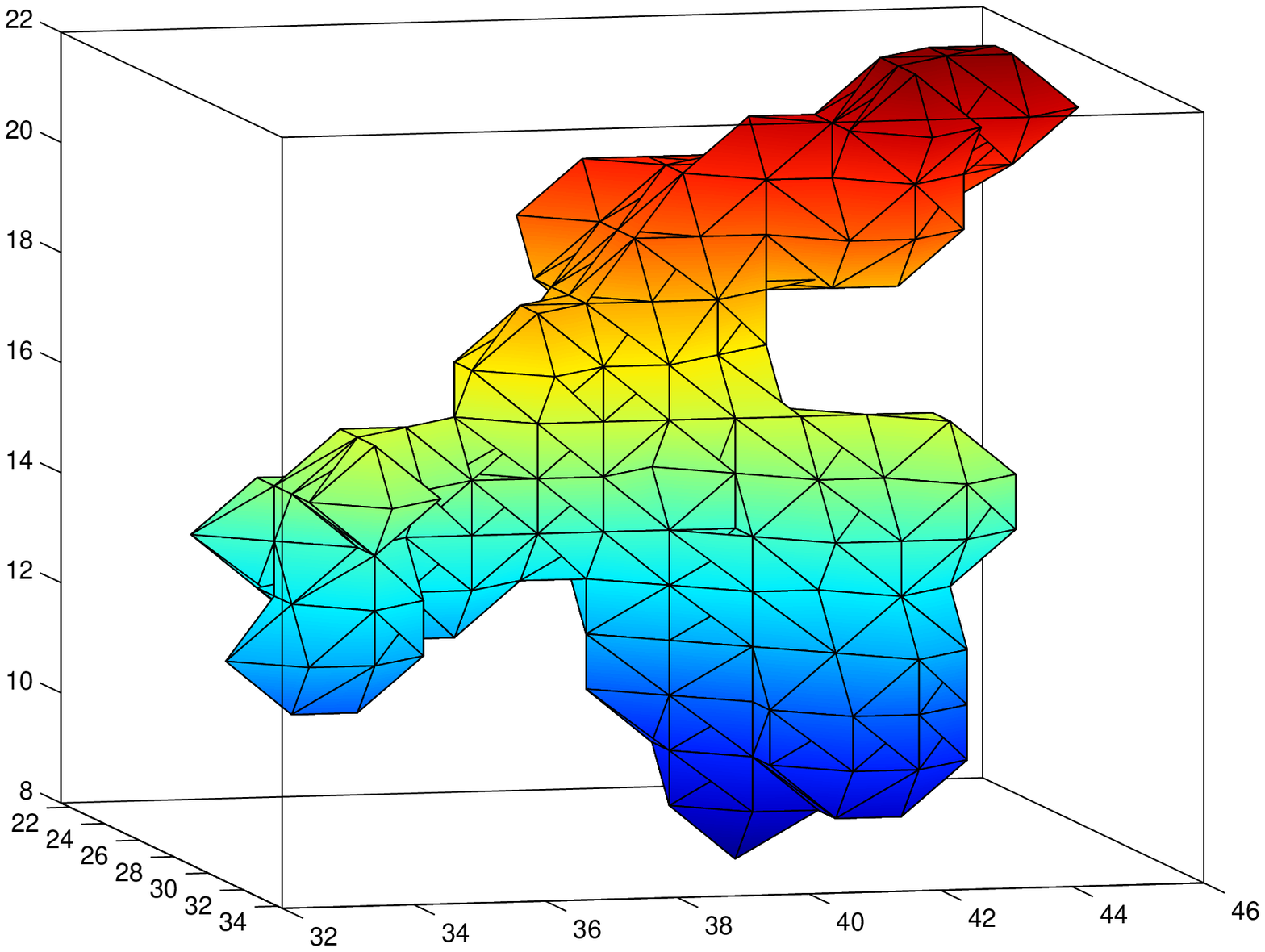}
\includegraphics[width=3in]{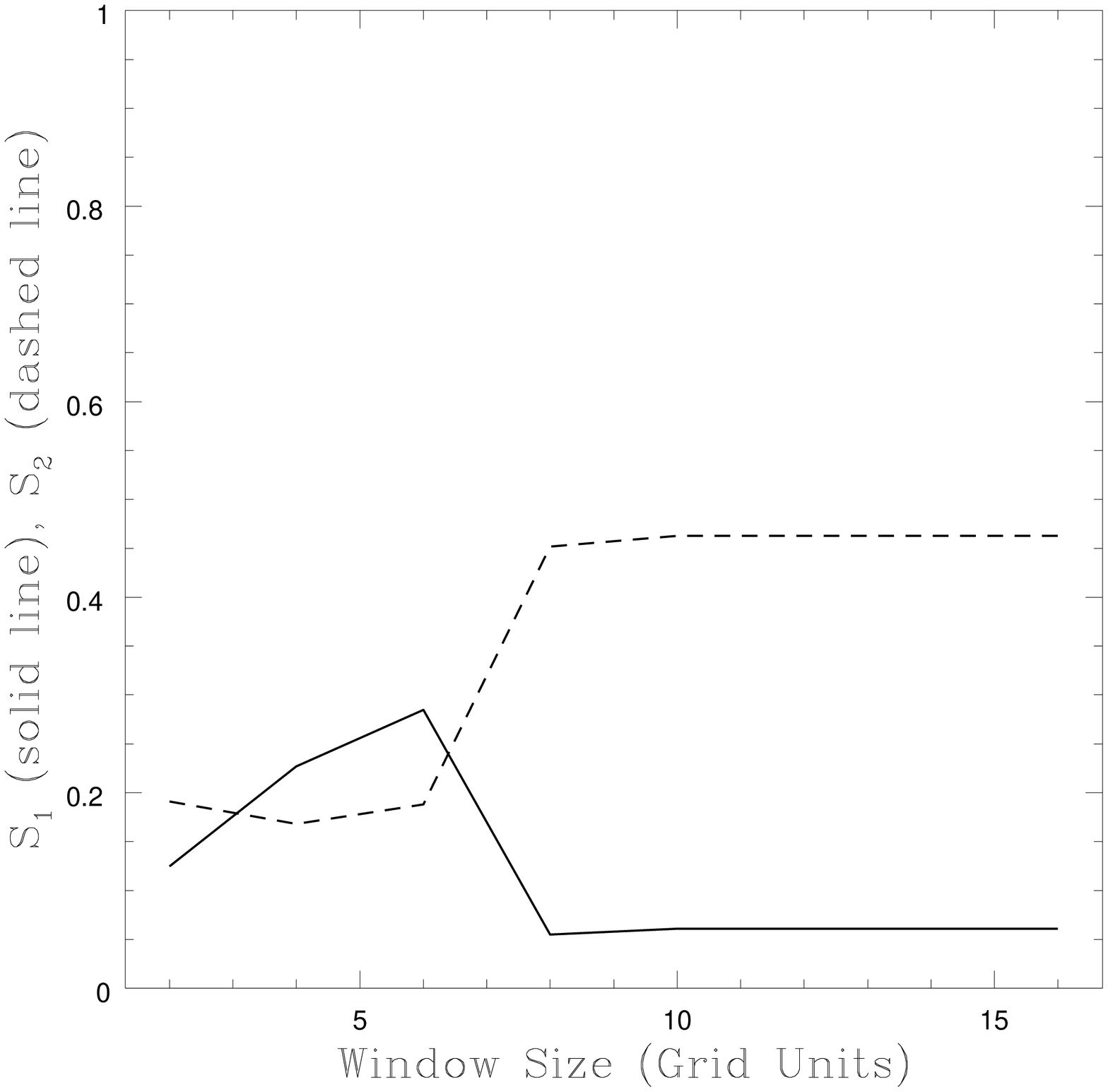}
\includegraphics[width=3in]{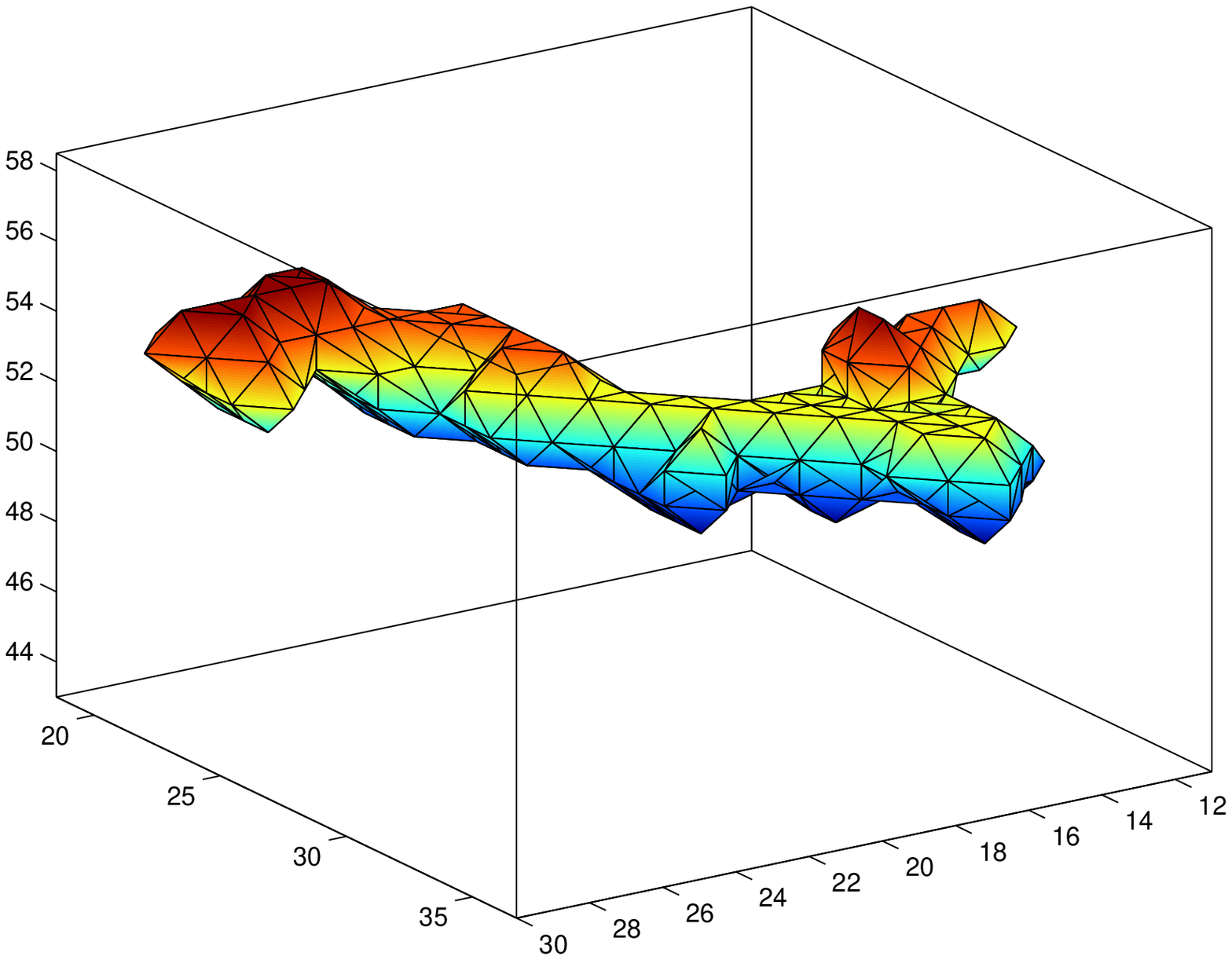}
\includegraphics[width=3in]{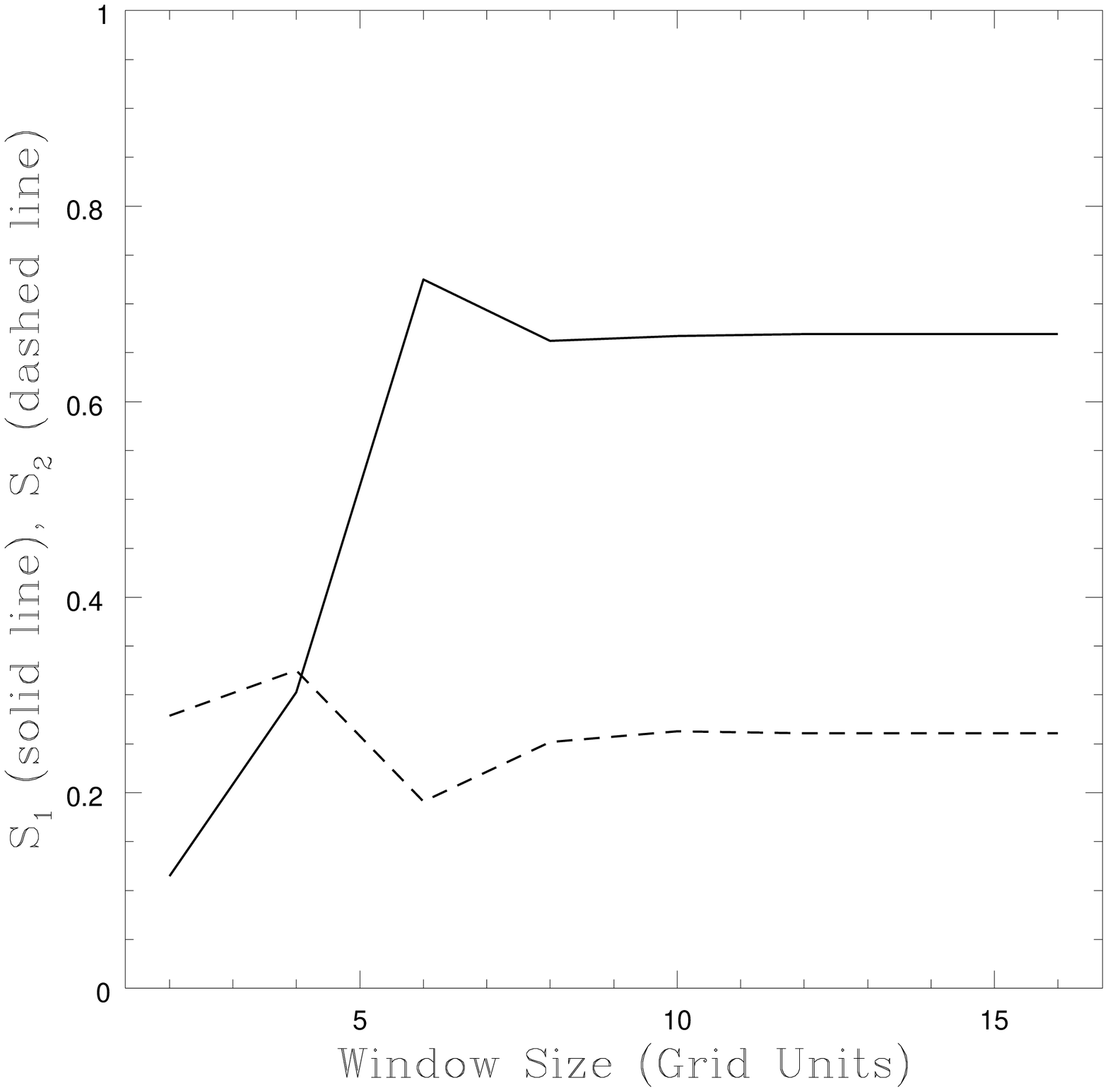}
\includegraphics[width=3in]{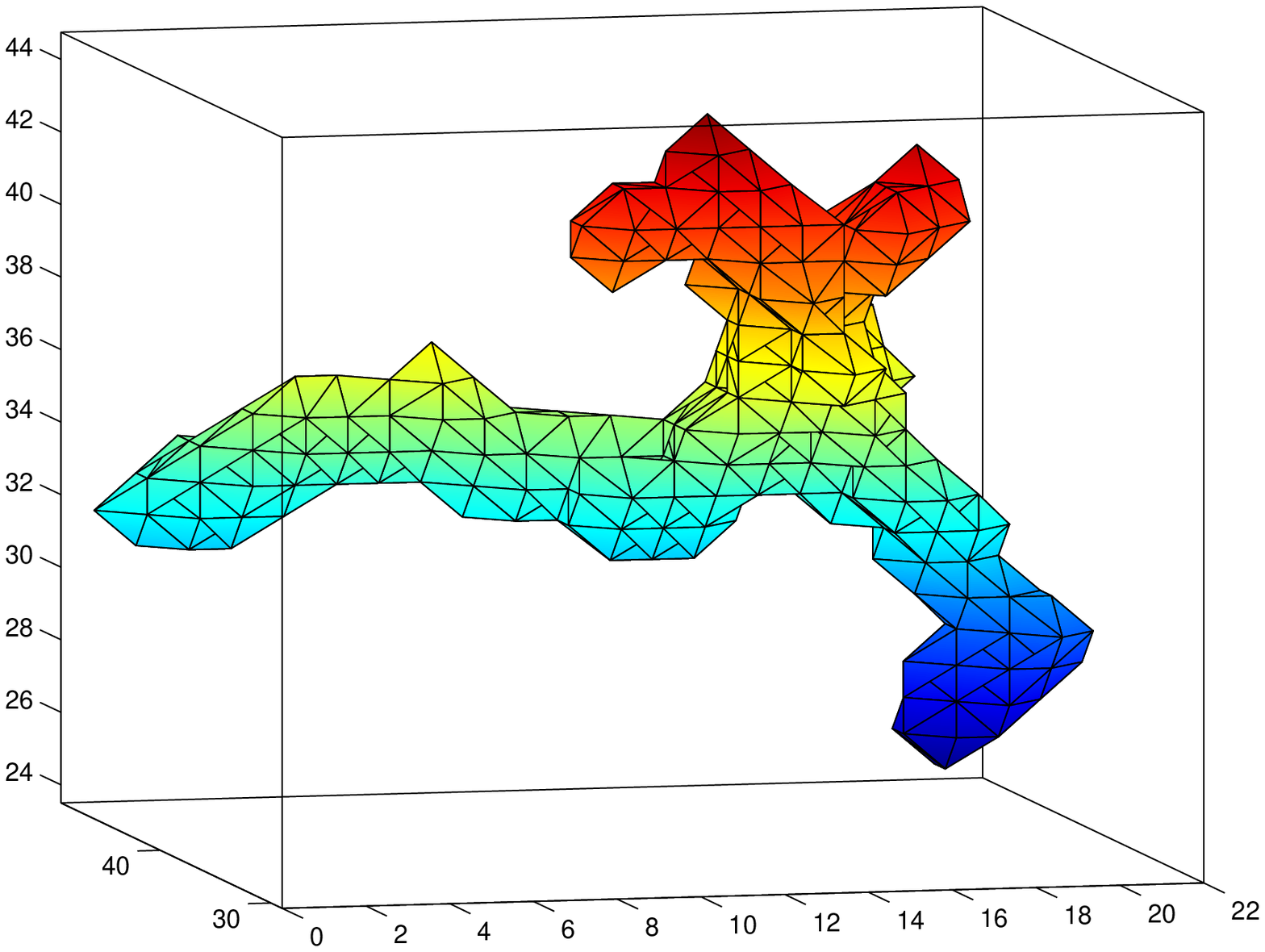}
\includegraphics[width=3in]{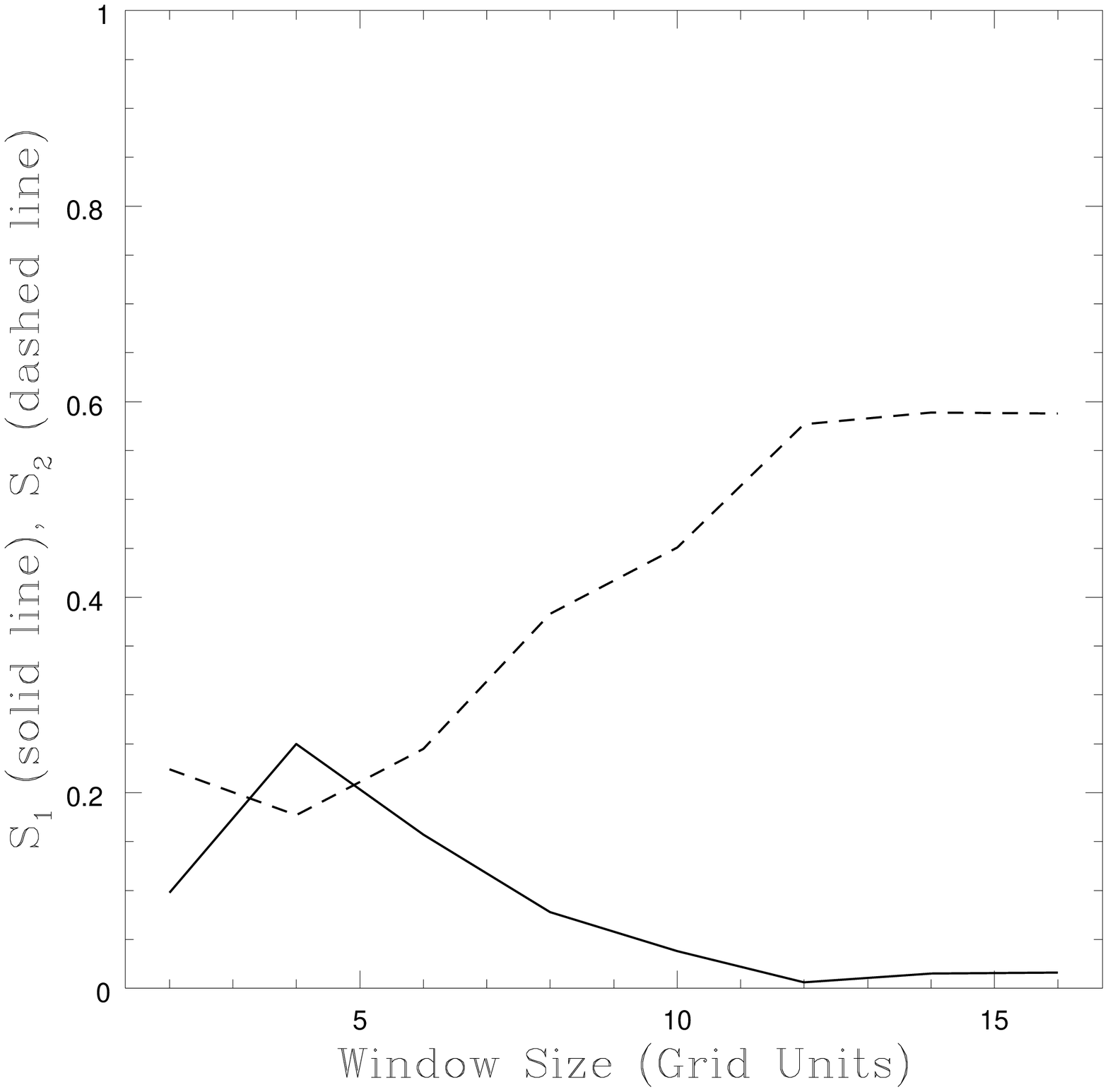}
\caption {Some clusters (left panels) occurring in our N-body simulations
are shown together with their BS \& LV shape statistics as 
functions of window size (right panels.}
\label {fig:sample.structures}
\end{figure}

In Fig.~\ref{fig:sample.structures} we have shown three carefully selected
objects chosen from our N-body simulations together with the BS 
statistics plotted next to them as a function of the window size. In 
each case the origin is chosen to be the centre of mass of the
object. In (a) we show a cluster which is more planar than it
is filamentary, in (b) a bent filament and in 
(c) we have a cluster with several tentacles emerging from it.
The shape functions plotted as a
function of the window size correctly reveal the morphological nature of 
these objects on small scales but on the scale of the object 
itself,
the shape revealed by the statistics can occasionally  be in conflict
with visual impression. In the case of (c) we clearly have a 
collection of several one-dimensional objects joined at the centre,
yet the statistics declare such an object to be planar. Perhaps this is
not entirely in correct since in as the number of 
tentacles begins to be large we do perceive the object to be planar. However,
the point is that even if we do have a fine filament, but bent, the object
would acquire some degree of planarity according to both BS and LV. However,
{\it shapefinders} based on Minkowski functionals are able to differentiate
between a wavy filament and a planar distribution.  (We would like to 
point out that the BS and LV statistics differ only in the value
of oblateness or planarity which they attribute to a given structure,
the measures of filamentarity $S_1$ and $S_l$ 
agree remarkably for all structures.) 

Our considerations in this Section have clearly shown that 
the LV statistic is not suitable for the description of morphology due to
its strong bias towards planarity. Consequently, in the rest of this paper
we restrict its application to fewer cases than the BS
statistics and rely on the latter to draw our main conclusions.

\section {Evolution of cluster shape in N-body Simulations}

We now apply the BS shape statistics to a systematic study of the evolution of
morphology of large scale structure in N-body simulations of gravitational
clustering.
(Details of our N-body simulations have been discussed in 
Section \ref{sec:nbody}) In the first part of this Section,
we present a statistical
analysis of our study of isolated clusters. There we discuss the
shape distribution function or shape-spectrum (a plot showing the number of
objects as a function of shape parameters) as well as the relationship between 
cluster mass and cluster shape.
In the second part, we discuss the morphology of
high density regions obtained by averaging shape
functions over a large sample of random but high density points.

\subsection {Morphology of isolated clusters in N-body simulations}

\begin {figure}
\centering
\begin {minipage}[c]{3in} \includegraphics[width=3in]{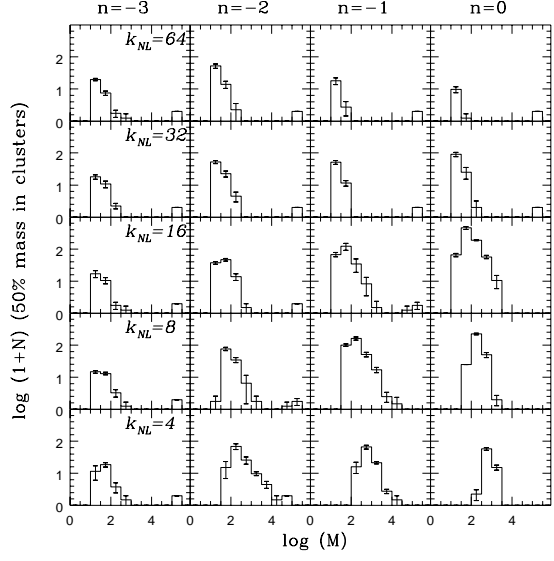} \end{minipage}
\caption {Multiplicity function: number of clusters as a function of 
cluster mass (arbitrary units) for clusters defined using 
the criteria that 50\% of the total mass is contained in clusters.}
\label {fig:multiplicity}
\end{figure}

\begin {figure}
\centering
\begin {minipage}[c]{3in} \includegraphics[width=3in]{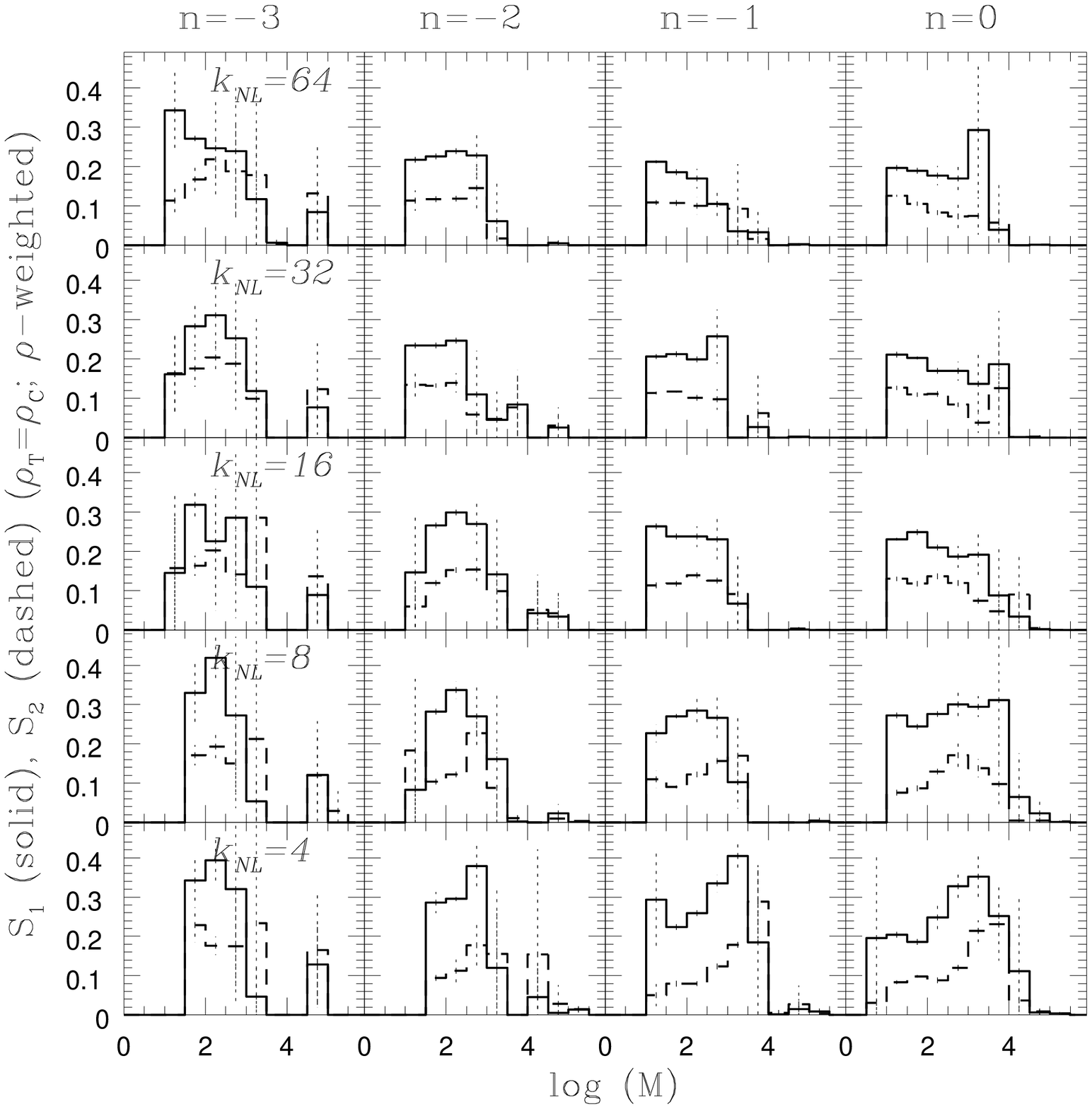}
\end {minipage}
\begin {minipage}[c]{3in} \includegraphics[width=3in]{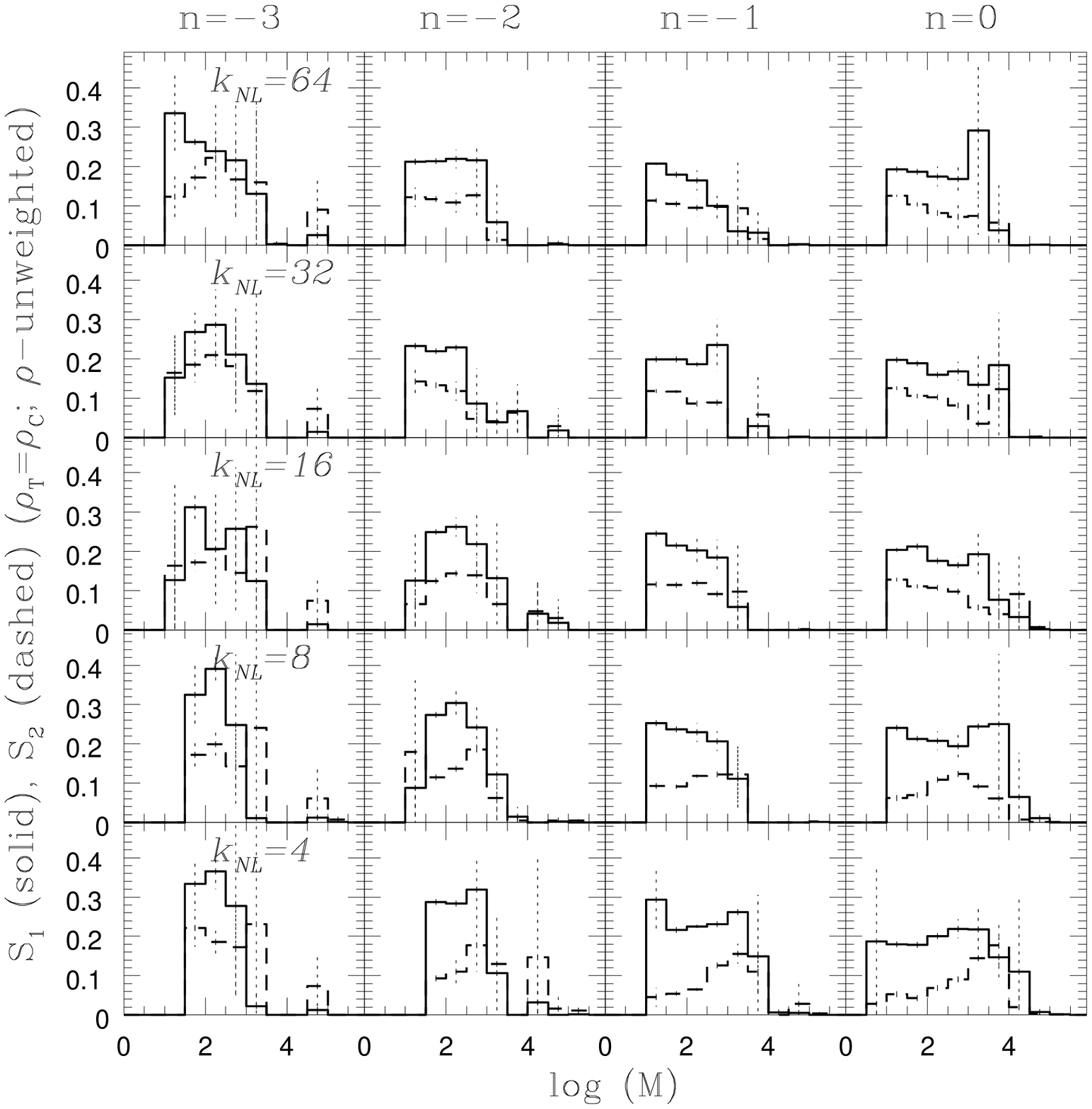}
\end {minipage}
\caption {Cluster shapes (BS) are shown as
functions of cluster mass. The
moments are density-weighted in the panels on the left and density-unweighted
in the panels on the right. (Clusters are defined at the percolation threshold.)}
\label {fig:shapes.bs.perc}
\end{figure}

We identify clusters in an N-body simulation using
two different methods and evaluate the shape parameters for each case.
The first method is based on percolation analysis briefly reviewed
in section \ref{sec:percolation} The second is more elementary:
we choose a density threshold such that 50\% of the total mass is
in high density regions. The histogram showing number of clusters obtained
using the second of these methods is plotted against cluster mass
in Fig.~\ref{fig:multiplicity}, for 
spectra with $n=-3,\ -2,\ -1,\ 0$ (left to right) and expansion
epochs $k_{\rm nl}=64,\ 32,\ 16,\ 8,\ 4$ (top to bottom).
(Recall that steeper spectra have more large scale power and larger
$k_{\rm nl}$'s correspond to earlier epochs.) 
Error bars in all plots (sometimes shown as dotted lines, eg. 
as in Fig. \ref {fig:shapes.bs.perc}) are computed using four
independent realisations of each spectrum.
For all spectra (with the possible exception of $n = -3$) we find that
most of the matter
gets transferred to larger clusters as the simulation evolves
(presumably due to merging). Very large superclusters present during
early stages are frequently not long lived objects but tend to disintegrate
in models with sufficient small scale power.  

After identifying clusters we evaluate the shape functions of each cluster
[($S_1, S_2$) for BS and ($S_l, S_p$) for LV] 
by choosing: (a) the origin to be the centre of mass
of the cluster and (b) a window large enough to enclose
the entire cluster.  We then ask essentially two questions: (i) What
is the {\em distribution function of shapes}? That is, how many clusters
are there of a given shape? We refer to the shape distribution function
as {\em shape-spectrum.} (ii) What is the relationship between cluster mass 
and its shape or, perhaps equivalently, cluster size and its shape.

\begin {figure}
\centering
\includegraphics [width=3 true in] {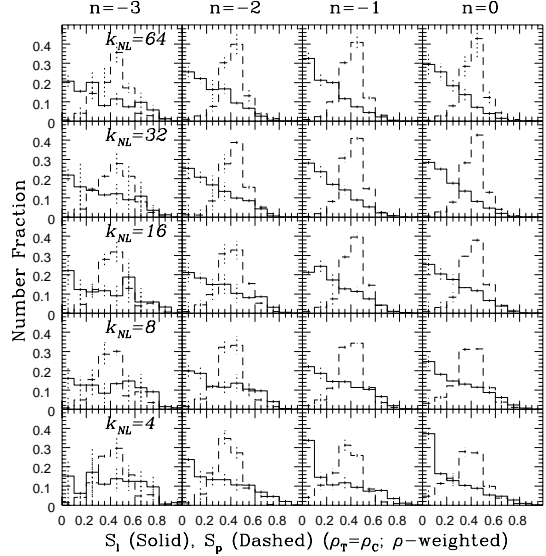}
\caption {Shape-spectrum for clusters using LV statistics. Clusters are defined at
the percolation threshold, the shape parameters are density-weighted.
The excess of clusters with $S_p \sim 0.4$, can be attributed to the bias of
LV towards planar structures demonstrated earlier in Fig.~\protect{\ref{fig:phase.space}}
and Fig.~\protect{\ref{fig:scatter}}.}
\label {fig:spectrum.lv.wt}
\end{figure}

\begin {figure}
\centering
\begin {minipage}[c]{3in} \includegraphics[width=3in]{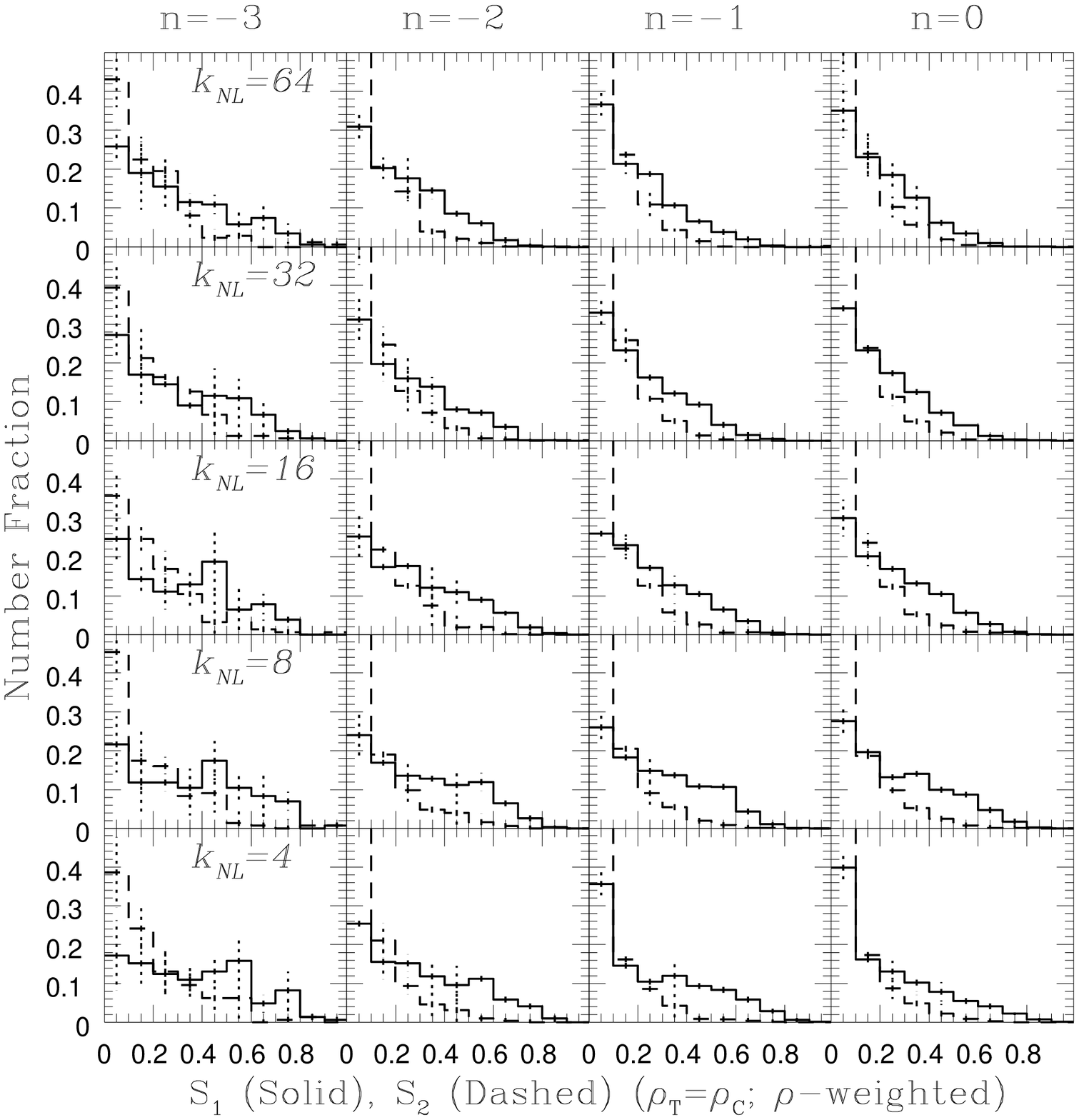}
\end {minipage}
\begin {minipage}[c]{3in} \includegraphics[width=3in]{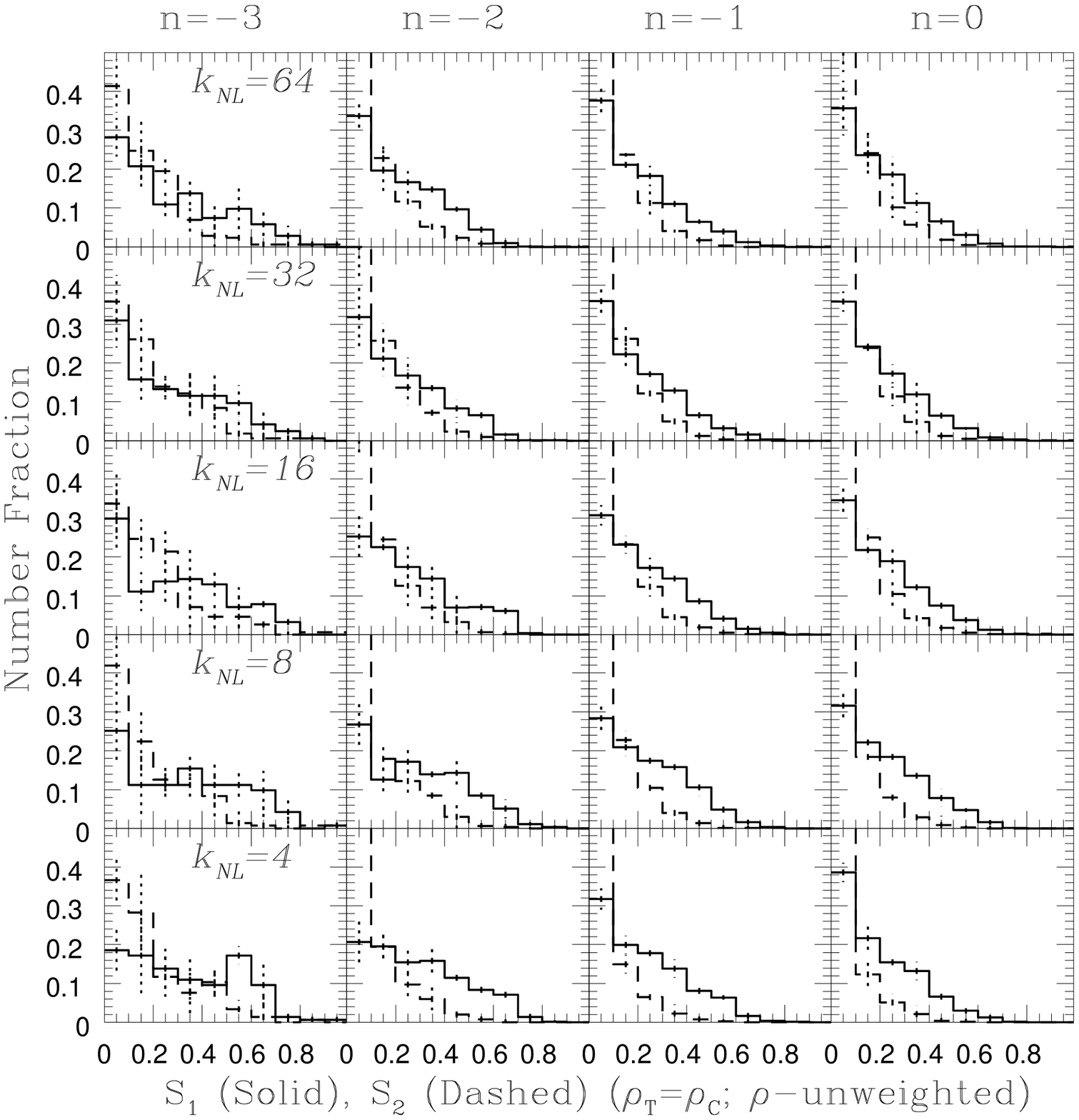}
\end {minipage}
\caption {Shape-spectrum for clusters using BS statistics. Clusters are defined 
at the percolation threshold; shape parameters are
density-weighted for panels on the left and unweighted for those on the
right. Note the moderate increase in filamentarity over pancakeness
which gets accentuated with epoch.}
\label {fig:spectrum.bs.perc}
\end{figure}

\begin {figure}
\centering
\includegraphics [width=3 true in]{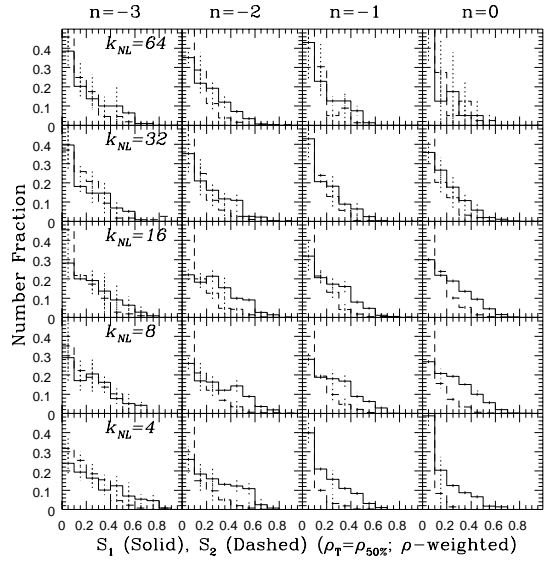}
\caption {Shape-spectrum for clusters using 
BS statistics. Clusters are defined using the criteria
that 50\% of the total mass is contained in clusters above the specified threshold.}
\label {fig:spectrum.bs.0.5wt}
\end{figure}

Fig.~\ref{fig:spectrum.lv.wt}, ~\ref{fig:spectrum.bs.perc} and 
\ref{fig:spectrum.bs.0.5wt} show the shape-spectrum of clusters in
various models and at different epochs of our N-body simulation.
Fig.~\ref{fig:spectrum.lv.wt} is for LV statistics, the rest
are for BS statistics. In Fig.~\ref{fig:spectrum.lv.wt} and
\ref{fig:spectrum.bs.perc} clusters are identified at the percolation threshold
and both density-weighted (Fig.~\ref{fig:spectrum.lv.wt}
and Fig.~\ref{fig:spectrum.bs.perc}~left)
and unweighted (Fig.~\ref{fig:spectrum.bs.perc}~right)
moments are constructed. 
In Fig.~\ref{fig:spectrum.bs.0.5wt} (as in Fig.~\ref{fig:multiplicity})
the density threshold for defining
clusters is chosen such that the cumulative mass in clusters equals
half the total mass in the simulation and the shape functions are obtained
using density-weighted moments.
(For both density-weighted and unweighted moments the threshold at which clusters
are defined is the same, the difference between the two cases arises because
in the former Eqs.~(\ref{eq:bs4}) and (\ref{eq:bs5}) are directly 
used to find the moments, whereas in the latter $\rho$ 
is set equal to unity in Eqs.~(\ref{eq:bs4}) and (\ref{eq:bs5}).)

\begin {figure}
\centering
\includegraphics [width=3 true in]{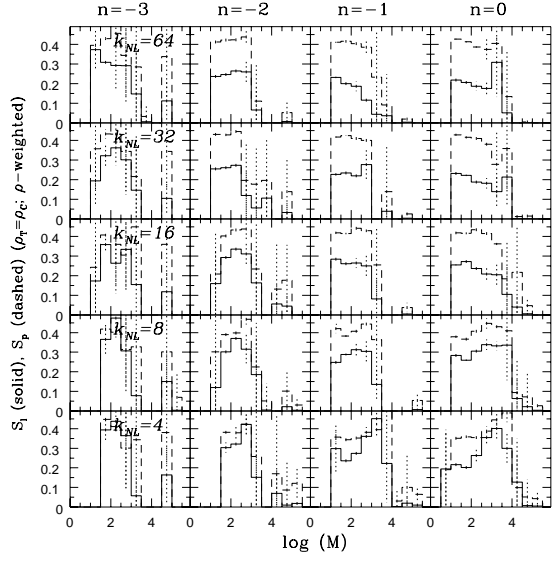}
\caption {Cluster shapes (LV) are shown as functions of cluster mass.
Clusters are defined at the percolation threshold.
Note the pronounced increase in 
filamentarity with cosmological epoch.}
\label {fig:shapes.lv.wt}
\end{figure}

\begin {figure}
\centering
\includegraphics [width=3 true in]{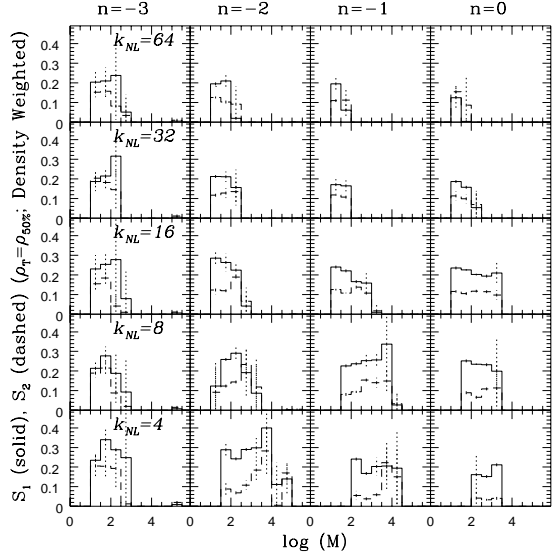}
\caption {Cluster shapes (BS)
are shown as functions
of cluster mass for clusters defined according to the criteria that 
50\% of the total mass is contained in clusters above the specified 
threshold.
Note the pronounced increase in 
filamentarity with cosmological epoch.}
\label {fig:shapes.bs.0.5wt}
\end{figure}

In Fig.~\ref{fig:shapes.lv.wt}, \ref{fig:shapes.bs.perc}  and 
\ref {fig:shapes.bs.0.5wt} (counterparts of 
Fig.~\ref{fig:spectrum.lv.wt}, \ref{fig:spectrum.bs.perc} and
Fig.~\ref{fig:spectrum.bs.0.5wt}, respectively) we plot
histograms of filamentarity and planarity for clusters 
of different masses.  Our results can be
summarised as follows: (a) The LV statistic, shown in 
Fig.~\ref{fig:shapes.lv.wt} consistently gives a larger
value for oblateness than BS. Thus, according to LV, most objects in 
N-body simulations have a larger degree of planarity which is very
prominent at the beginning. However, filamentarity grows faster than
planarity as gravitational clustering advances, and, at the last epoch
both planarity and filamentarity become almost equal.
The study of previous Sections suggests that the excess of planarity over 
filamentarity is due to the manner in which the LV statistic is 
constructed and should not be regarded as a real physical effect. For this
reason we attribute greater significance to the results of the BS
statistic. 
The results of applying the BS statistic to clusters in N-body
simulations are shown in Fig.~\ref{fig:shapes.bs.perc}  and 
Fig.~\ref{fig:shapes.bs.0.5wt}. We note that the
result of density weighting the moments has the effect of enhancing
the shape signals, especially those of high mass clusters and during
late stages of gravitational instability (this can be
seen by comparing the bottom-most two rows of panels in 
Fig.~\ref{fig:shapes.bs.perc}). 
We feel this to be  indicative of the fact that {\em high density regions of 
massive clusters/superclusters tend to be more filamentary or planar/ribbon-like}. 
The advantage of using a fixed mass fraction to identify clusters is that
we essentially follow a fixed fraction of the total mass as
gravity binds objects into larger and larger structures. Thus, the plots in
Fig.~\ref{fig:shapes.bs.0.5wt} show how matter
distributed homogeneously (top panels) to begin with,
clumps together
to form definite shapes, as evident from dramatic increase in the
amplitudes of $S_1$ and $S_2,$ finally ending up in giant filaments/pancakes
as the simulation evolves.\footnote{We would like to point out that objects
described as `pancakes' or `filaments'
could in fact equally well be ribbons, since,
as demonstrated in Table \ref{table:tor}
neither BS nor LV is equipped to differentiate
between ribbons and pancakes, or for that matter between ribbons and
filaments. This issue will be discussed in greater detail in
Buchert et al. 1998.}
We would also like to point out that the histograms
of filamentarity $S_l$ and $S_1$ in Fig.~\ref{fig:shapes.lv.wt}
and \ref{fig:shapes.bs.perc} (left), are very nearly the same
(this can easily be seen by comparing corresponding panels). 
($S_1$ \& $S_l$ describe filamentarity in BS and LV respectively.) Thus,
$S_l$ gives virtually the same results as $S_1$ although $S_p$
is much larger than $S_2$ due to the former's greater bias towards 
planarity.

These figures clearly demonstrate that both
filamentarity and planarity are statistically significant at all epochs
of gravitational clustering and for all spectra analysed by us.
However, it is also true that filamentarity exceeds
planarity and the former grows more pronounced as clustering
evolves. Thus, the increase of filamentarity with epoch appears to be a
generic feature of gravitational clustering demonstrated 
by both LV and BS statistics and 
confirming earlier results by Sathyaprakash, Sahni
\& Shandarin (1996). 
(One should also point out that for spectra
with significant large scale power ($n = -3, -2$) the very largest structures
tend to have equal measures of planarity and filamentarity although
neither is large enough to be statistically significant.) We expect that 
future large redshift surveys, such as the 2 Degree Field (2dF) and 
Sloan Digital Sky Survey (SDSS), will 
resolve several important issues including: 
the distribution of filaments, pancakes and ribbons
in the Universe;
whether `Great Walls' seen in CfA and
SSRS surveys are genuine filaments or are merely perceived to be such
because we are viewing slices of a two-dimensional cellular network etc.
An analysis of the IRAS 1.2 Jy catalogue discussed in Sathyaprakash, Sahni, Shandarin \& Fisher (1998)
hints at some general trends and expectations.

\subsection {Average Shape Functions}

\begin {figure}
\centering
\includegraphics[width=3true in]{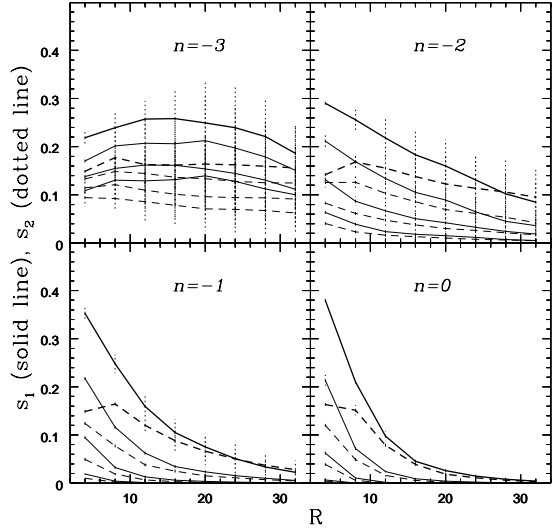}
\caption{BS shape parameters $S_1$, $S_2$ characterising filamentarity and 
planarity respectively, are shown as functions of the 
window size $R$ after averaging over a number of random but high 
density points. Only those points where the density was larger than the
percolation threshold were used as sample points around which to place
the window.  We clearly see the growth of filamentarity as clustering 
advances. Planarity also increases, but its growth is less pronounced.}
\label{fig:average}
\end{figure}

In this Section we discuss the morphology of 
typical high density regions in our N-body simulations. To assess this in
a meaningful manner we choose a number of random 
points lying above 
the percolation threshold.
We then evaluate BS shape parameters ($S_1, S_2$) as 
functions of radius of the window centered at each of these points and
then compute the average shape parameters and dispersion around that
average. Results obtained are shown in Fig.~\ref{fig:average}
where filamentarity (solid lines) and planarity (dashed lines)
are both plotted as functions of window size for various models
($n=-3,\ -2,\ -1,\ 0$) with heavier lines representing later
epochs. This graph shows a clear distinction between steeper and flatter
spectra. The increase of both filamentarity and planarity is spectacularly 
demonstrated and we find objects on larger scales growing
noticeably more prolate/oblate as 
the scale of nonlinearity advances.

\section {CONCLUSIONS AND FUTURE DIRECTIONS}

In this paper we have addressed the issue of morphology of clusters and
superclusters in N-body simulations of gravitational clustering. 
We have made an exhaustive comparison of two shape statistics proposed
respectively by Babul \& Starkman (1992) and by Luo \& Vishniac (1995). We 
find that the LV statistic is biased towards oblate structures and consistently
predicts a larger value for this property than BS.  
Being moment-based statistics, both BS and LV have difficulties in describing 
curved and topologically nontrivial objects which arise at
lower density thresholds in N-body simulations and in fully 
three-dimensional catalogues of large scale structure. 
We therefore conclude that 
moment-based shape statistics provide only a 
partially correct description of morphology and must be complemented 
by other shape indicators, such as Shapefinders, which are derived from 
Minkowski functionals and are, therefore, not moment-based (\cite{sss98}). 

Applying LV and BS to an analysis of cluster/supercluster shapes in 
N-body simulations of gravitational clustering we find a marked increase
in both filamentarity and planarity, the former becoming more
pronounced as the simulation evolves. This result is true for all spectra
considered by us and for all epochs.
Thus, matter gets collected into objects of ever increasing size
having pronouncedly filamentary/pancake/ribbon-like properties.
Upcoming large redshift surveys, such as the Sloan Digital Sky Survey (SDSS)
and the 2 Degree Field (2dF), combined with larger
simulations of gravitational clustering than the ones we consider here,
are bound to shed light on the morphology of the largest superclusters/voids in the Universe and on whether they can sensibly arise in models of gravitational
clustering. Some preliminary work in this direction will be reported in
Sathyaprakash, et al. (1998a), Sathyaprakash, et al. (1998b), 
Buchert et al. (1998).

\section* {Acknowledgments}

S. Shandarin acknowledges the support of NASA grant NAG 5-4039 and 
EPSCoR 1998 grant.

\clearpage

\begin{table}
\caption {LV $(S_l,~S_p)$ and BS $(S_1,~S_2,~S_3)$ shape
statistics for elliptical tori
of different dimensions together with the values obtained
by {\em Shapefinders} $({\cal K}_f,~{\cal K}_p)$. 
The second column (M) gives the
morphology of the object: P= Pancake, 
F= Filament, R= Ribbon, S= Sphere.  Note that while Shapefinders can distinguish 
filaments from ribbons and ribbons from pancakes LV and BS statistics cannot.} 
\begin {center}
\begin{tabular}{ccccc}
\hline
$(a,~b,~c)$ & M &  $(S_l,~S_p)$ & $(S_1,~S_2,~S_3)$ & $({\cal K}_f,~{\cal K}_p)$\\
\hline
$(15,~16,~~1)$ & P1 & $(0.01,~0.16)$ & $(0.00,~0.06,~0.87)$ & $(0.13,~0.78)$\\
$(~1,~16,~15)$ & P2 & $(0.00,~1.00)$ & $(0.00,~1.00,~0.00)$ & $(0.14,~0.78)$\\
$(~2,~30,~~2)$ & F  & $(0.00,~0.99)$ & $(0.00,~0.99,~0.06)$ & $(0.84,~0.14)$\\
$(~4,~28,~~1)$ & R1 & $(0.00,~0.97)$ & $(0.00,~0.96,~0.15)$ & $(0.78,~0.43)$\\
$(~1,~28,~~4)$ & R2 & $(0.00,~1.00)$ & $(0.00,~1.00,~0.00)$ & $(0.78,~0.43)$\\
$(15,~16,~15)$ & S  & $(0.01,~0.38)$ & $(0.00,~0.21,~0.71)$ & $(-0.08,~0.14)$\\
\hline
\end{tabular}
\end {center}
\label{table:tor}
\end{table}

\begin {figure}
\centering
\includegraphics [width=3 true in] {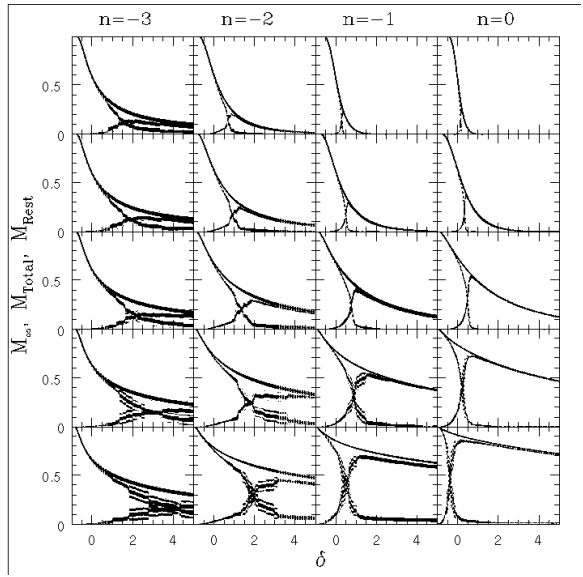}
\caption {Percolation curves are plotted as functions of the density contrast
$\delta$, for scale invariant N-body simulations
with different initial spectra (left to right) and at increasing cosmological  
epochs (top to bottom). Shown are (a) $M_\infty$ -- the mass fraction in the 
largest cluster (dotted), (b) $M_{\rm Rest}$ -- mass fraction in all clusters with the exception of
the largest (dashed) and (c) $M_{\rm Total}$ -- mass fraction in {\it all} clusters above a 
density threshold (solid), $M_{\rm Total} = M_\infty + M_{\rm Rest}$.
The percolation transition is marked by a sharp
increase in $M_\infty$.}
\label {fig:perc}
\end{figure}

\end{document}